\documentclass[12pt,preprint]{aastex}
\defcitealias{burles:98a}{}
\defcitealias{burles:98b}{}
\defcitealias{crighton:04}{}
\defcitealias{hebrard:03}{}
\defcitealias{kirkman:03}{}
\defcitealias{omeara:01}{}
\defcitealias{pettini:01}{}
\defcitealias{sembach:04}{}
\defcitealias{centurion:03}{}

\begin{document}

\title {Cosmic Star Formation, Reionization, and \\
Constraints on Global
Chemical Evolution}

\author{Fr\'{e}d\'{e}ric~Daigne}
\affil{Institut d'Astrophysique, 98 bis Boulevard Arago, Paris 75014, France\\
and Universit\'{e} Pierre et Marie Curie -- Paris VI, 4 place Jussieu, 75005 Paris, France}
\email{daigne@iap.fr} 

\author{Keith~A.~Olive}
\affil{ William I. Fine Theoretical Physics Institute, School of Physics and Astronomy, \\
University of Minnesota, Minneapolis, MN 55455 USA}

\author{Elisabeth~Vangioni-Flam}
\affil{Institut d'Astrophysique, 98 bis Boulevard Arago, Paris 75014, France}

\author{Joseph~Silk}\affil{Department of Physics, University of Oxford, Keble Road, Oxford OX1 3RH, \\
and Institut d'Astrophysique, 98 bis Boulevard Arago, Paris 75014, France}
\and  

\author{Jean~Audouze}
\affil{Institut d'Astrophysique, 98 bis Boulevard Arago, Paris 75014, France}

\begin{abstract}
%

Motivated by the WMAP results indicating an early epoch of
reionization, we consider alternative  cosmic star formation models which
are capable of reionizing the early intergalactic medium.  We develop
models which include an early burst of massive stars (with several possible
mass ranges) combined with standard star formation.  We compute the stellar ionizing flux
of photons and we track the nucleosynthetic yields for several
elements\,: D, $^{4}$He, C, N, O, Si, S, Fe, Zn.  We compute the
subsequent chemical evolution as a function of redshift, both in the
intergalactic medium and in the interstellar medium of
forming galaxies,
 starting with  the primordial objects  which are
responsible for the reionization.  We apply constraints from the
observed abundances in the Lyman $\alpha$ forest and in Damped Lyman
$\alpha$ clouds in conjunction with the ability of the models
to produce the required degree of reionization.  We also consider
possible constraints associated with the observations of the two
extremely metal-poor stars HE 0107-5240 and CS22949-037.  We confirm
that an early top-heavy stellar component is required, as a standard star
formation model is unable to reionize the early Universe and reproduce
the abundances of the very metal-poor halo stars.  A bimodal (or
top-heavy) IMF (40 - 100 M$_\odot$) is our preferred scenario compared
to the extreme mass range ($\ga$ 100 M$_\odot$) often assumed to be
responsible for the early stages of reionization.
A mode of  even more extreme stellar masses in the  range ($\ge$ 270 M$_{\odot}$) has
also been considered. All massive stars in this mode collapse entirely into
black holes, and as a consequence, chemical evolution and reionization
are de-correlated. The ionizing flux from these very massive stars can
easily reionize the Universe at $z\sim 17$. However the chemical
evolution in this case is exactly the same as in the standard star
formation model, and the observed high redshift abundances are not reproduced.
We show that the initial top-heavy mode, which
originally was introduced to reionize the early Universe, produces 
rapid initial metal pollution. The existence of old, C-rich halo stars
with high [O/Fe] and [C/Fe] ratios is predicted as a consequence of
these massive stars. The recently observed abundances in the oldest
halo stars could trace this very specific stellar population. The extreme mass range is disfavored and 
there is no evidence, nor any need, for a hypothesised 
primordial population of  very 
massive 
stars in order to account  for  the 
chemical abundances of extremely metal-poor halo stars
or of the intergalactic medium.

The combined population of early-forming, 
normal (0.1 - 100 M$_\odot$) and massive (40 - 100 M$_\odot$) stars
can simultaneously explain 
the cosmic chemical evolution and the observations of
extremely metal-poor halo stars
and also account for early 
cosmological reionization.

\end{abstract}


\section{Introduction}

With the recent release of WMAP data, we have without question,
entered the age of precision cosmology.  While many of the accurately
determined cosmological parameters were found to have values
consistent with a $\Lambda$CDM cosmology, the $TE$ cross-correlation
power spectrum is consistent with a large optical depth due to
electron scattering, $\tau_e = 0.17 \pm 0.04$ \citep{kogut:03} which suggests a very
early epoch ($z \simeq 20$) of reionization in the inter-galactic
medium (IGM). It is often suggested that reionization is due to a
population of very massive stars forming in minihalos, cooled
predominantly by H$_2$ \citep[see][]{cen:03a,haiman:03,wyithe:03,bromm:04}. It has also been
suggested that this early period of reionization is not total, but
rather is regulated by radiative feedback effects delaying the
complete reionization to a later redshift ($z \simeq 6$) accounting
for the observed Gunn-Peterson trough \citep{becker:01}.

Even a relatively brief period of massive star formation at high
redshift  can have dramatic consequences on the chemical enrichment of
the primitive structures and of the IGM.  There has been significant progress in our
understanding of the yields of massive stars, particularly for
so-called pair-instability supernovae, allowing one to trace the
chemical history as influenced by an early massive population of
stars.  If indeed the first generation of stars were very massive
(140-260 M$_\odot$), they would act as prompt initial enrichment
sources for
the IGM as well as the progenitors of galaxies such as our own.  In
addition, they would lay down a very distinct chemical signature which
can be compared with other potential initial mass functions for the
first generation of stars capable of producing the required degree of
reionization.

The argument for a prompt initial enrichment laying down the seeds for
the chemical evolution of the Galaxy is an old one going back to
\citet*{truran:71}.  This notion receives confirmation in
observations of r-process elements as a function of [Fe/H] in halo
stars \citep*{wasserburg:00,qian:01}.  The
question of the ionization capacity of the massive stars supposed to
produce this initial enrichment was recently considered by 
\citet{oh:01}, \citet{oh:02}, and \citet*{venkatesan:03b}
in the context of a late period of reionization at  a redshift $z \sim 6$.
More recently, metallicity studies of Damped Lyman
$\alpha$ clouds (DLAs) at high redshift
also indicate the presence of an initial metallicity of roughly [Fe/H]
$\sim 10^{-3}$ \citep{prochaska:03}.  \citet*{tumlinson:04}
examined this question in the context of an early period of reionization 
as indicated by the WMAP data and present a general discussion of the 
initial mass function (IMF) of the first stars and their associated 
nucleosynthesis. Specifically they use the measured element abundances in
the metal-poor halo stars to derive the IMF of population III stars responsible 
for the reionization of the early IGM. While our goal here is similar, we 
include a comprehensive model of chemical evolution to address this question.  

Related to the question of an early population (Pop III) generation of
stars is the observation of extremely metal-poor ([Fe/H] $<$ -4) stars such as
HE 0107-5240 (\citealt{christlieb:04} and \citealt*{bessell:04}), 
a halo giant which is extremely
metal-poor, [Fe/H] = -5.3.  While the abundances of other observed
elements, e.g., Na, Mg, Ca, etc. are (roughly) as low, the abundance
of C (and N and O) is surprisingly high, [C/H] $\approx -1.3$ ([N/H]
$\approx -3.0$, [O/H] $\approx -2.9$).  Indeed the very existence of
this star poses a challenge for both models of chemical evolution (and
also stellar evolution) and star formation as its mass is believed to
be quite low ($M \sim 0.8$ M$_\odot$).  We note, however, that it is
possible that this star (and other similar stars such as CS 22949-037,
\citealt{depagne:02} and \citealt{israelian:04}) was born under very particular circumstances as
these abundance patterns may be explained by the pre-enrichment of a
massive zero-metallicity star (with fall-back) 
(\citealt*{umeda:03}; see also \citealt{suda:04}),
or perhaps the star was polluted by a binary companion in its AGB
phase.  Thus with some degree of caution, we will discuss the
implications of the existence of this star with respect to the models
considered.

While there have been several detailed studies concerning the
ionization efficiency of massive stars, attention to chemical
enrichment has focused on the overall metallicity produced in
structures with massive or very massive stars.  Here, we calculate the
combined effect of an early population of massive stars on the
reionization history and chemical enrichment history of the early cosmic
structures and of the IGM. In the framework of the hierarchical galaxy 
formation model, most of the halo, and in particular a DLA, is built
up by smaller merging systems that are tidally disrupted (see e.g. \citep{zentner:03}).
Therefore, they inherit the metals produced at large z by population III stars 
as a Prompt Initial Enrichment (PIE). The intensity of this PIE must
be limited to avoid a metallicity surpluss in these late structures which produce 
new metals via the normal mode of star formation. Thus our approach is the following :
(i) since the metallicity increases very rapidly as soon as the first population III 
stars explode, the timescale of the massive mode of star formation is necessarily short
(see section 4.3); (ii) therefore, the cosmic SFR at $z \la 5-6$ reflects only the normal mode of star formation. This allows one to normalize its intensity to fit the observed cosmic SFR at these
redshifts ; (iii) once this normalization has been done, the metal production by the normal
mode is immediately derived without any other parameter adjustement; (iv) from the amount of metals
produced by the normal mode, it is then possible to estimate the maximum permitted initial 
metallicity in the structures at the beginning of the normal mode to avoid the over-production
of metals  in the 
late structures; (v) finally, this allows us to normalize the intensity of the early massive mode (pop. III stars) as it governs the PIE. 
We consider several initial mass functions
which differ primarily in the population of massive stars forming at
zero (or near zero) metallicity.  Given an initial mass function and
an associated star formation rate, we calculate in a consistent way 
both the chemical history of the cosmic structures and of the IGM as well as the
efficiency for its reionization. Compared to previous studies, we do not 
restrict our discussion to the global metallicity but we specifically 
follow several chemical elements (D, $^{4}$He, C, N, O, Si, S, Fe, Zn).
We further discuss the implications
of the observation of the extremely metal poor star ([Fe/H] = -5.3),
HE 0107-5240.  This star can severely constrain such models and/or
models of stellar nucleosynthesis, particularly those for extremely
massive stars which evolve through pair instability.

We find that a bimodal (or top-heavy) IMF (40 - 100 M$_\odot$),
for example as discussed by \citet{ciardi:03},
 is
preferred over a scenario which includes a component with very massive
stars ($\ga$ 100 M$_\odot$) which are often assumed to be responsible
for the early stage of reionization.  The best model is well suited to
account for C and Zn abundances, as well as Fe, O, Si and S in the
DLAs, in the metal-poor IGM, and in the oldest halo stars.
Furthermore, in order to reionize the universe at $z\sim 17$, we
require an escape fraction for ionizing photons of only about 5 \%.
In contrast, we find that the massive starburst model, relying mostly
on pair instability supernovae (PISNae) from stars in the range 140--260
M$_{\odot}$, requires a lower SFR intensity due to the overproduction
of metals and as a consequence, the UV flux is weaker and requires a
higher escape fraction of ionizing photons ($\sim$ 30 \%) to achieve
early reionization. Note that \citet{venkatesan:03b} have also found a low ionizing efficiency for
very massive stars.  Nevertheless, the overproduction of some
$\alpha$--elements (specifically S and Si) seems unavoidable in these
models.  Moreover due to the specific yields of these PISNae, the
observations of the metal-poor halo stars are not reproduced.  We also
consider a model with an extremely massive component, i.e., with a
stellar mass range $\ge$ 270 M$_{\odot}$. Assuming that these massive
stars collapse entirely into black holes in this mode, chemical
evolution and reionization are de-correlated. We find that the
ionizing flux from these very massive stars can very easily reionize
the Universe at $z\sim 17$. However we also find that the resulting
chemical evolution in this model is exactly the same as in a standard
star formation model, and the high redshift abundances are not
reproduced.

The outline of the paper is as follows: in section 2, we describe the
chemical evolution models in detail, and our method for calculating
the chemical enrichment inside the structures and the IGM.  In section
3, we describe our computation of the flux of ionizing photons. Our
results are collected in section 4 and our conclusions are summarized
in section 5.

\section{Chemical Evolution Models}
 
 While observations of stellar mass distributions are able to fix the
 present-day IMF quite well, one is required to apply observations of
 a wide range of element abundances in order to constrain the combined
 star formation rate and IMF throughout the history of Galaxy.  When
 one goes beyond galactic chemical evolution, recent observations of
 the cosmic star formation rate (through the cosmic luminosity
 function) \citep[][and references therein]{lanzetta:02,nagamine:03}, and the abundances of heavy elements
 in quasar absorption systems enable one to place constraints on
 models up to a redshift of a few \citep{pettini:02,pettini:03,ledoux:03,prochaska:03,centurion:03}.  
In contrast, there have
 been very few constraints on the very first epoch of star formation
 concerning the IMF and SFR.  The recent discovery of a candidate
 $z=10$ galaxy \citep{pello:04,ricotti:04a}
 indicates that this
 observational situation may improve rapidly.
 
 As noted above, one of the most startling results of the 1st-year
 WMAP analysis is related to the ionization history of the Universe.
 It is often claimed that achieving a high degree of ionization at a
 redshift $z \simeq 20$, requires a very early generation of massive
 stars.  A burst of massive star formation would have rather dramatic
 effects on the chemical history of the Universe.  Not only would such
 a population of stars lay down an initial metal enrichment for the
 IGM, they would, depending on the IMF, lay down a very distinct
 fingerprint based on the abundances of elements produced by these
 stars.  In particular, one could expect that the element abundance
 patterns determined by very massive stars ($\ga 100$ M$_\odot$) would
 look very different from that of a normal IMF or even one biased by
 massive stars with masses in the typical range 10 - 100 M$_\odot$.

To test this hypothesis, we will consider several types of chemical
evolution models.  A basic review of chemical evolution models can be
found in \citet{tinsley:80}.  In this paper, we present an extension of
this standard formalism, where we now take into account two gas
reservoirs, one accounting for the IGM and one for the interstellar medium (ISM) of the
forming galaxies. This allows one to obtain high-redshift abundances,
measured both in DLAs and in the Lyman-$\alpha$ forest. Galaxies form
hierarchically, and the building blocks are defined by the fact that
baryons must be able to cool in the dark matter halos. The minimum
scale is about $10^6\ \rm M_\odot,$ and these objects first condense at
about $z\sim 20.$ However for the purposes of chemical evolution and
reionization, we can avoid any detailed discussion of the mass
function by restricting ourselves to integrated quantities such as
condensed baryonic mass and the gas accretion rate,
and to  key parameters such as the escape
fraction of ionizing photons, the star formation efficiency and the chemical yields.

Each model will be defined by several parameters : (i) the initial
redshift at which star formation begins, $z_\mathrm{init}$. We adopt
$t=0$ at this redshift and relate the age at redshift $z$ of a star
formed at $z=z_\mathrm{init}$ by
\begin{displaymath}
\frac{dt}{dz} = \frac{9.78\ h^{-1}\
\mathrm{Gyr}}{\left(1+z\right)\sqrt{\Omega_\mathrm{\Lambda}+\Omega_\mathrm{m}(1+z)^{3}}}\ ,
\end{displaymath}
where we adopt $\Omega_\mathrm{\Lambda}=0.73$,
$\Omega_\mathrm{m}=0.27$ and $h=0.71$ \citep{spergel:03}. Notice
that the quantity plotted in the figures in this paper is not the time
$t$ but the age of the Universe, which is $t+t_\mathrm{init}$, where
the constant $t_\mathrm{init}$ is the age of the Universe at
$z=z_\mathrm{init}$.  (ii) the star formation rate (SFR), $\psi(t)$ ;
(iii) the initial mass function (IMF), $\phi(m)$, for $m_\mathrm{inf}
\le m \le m_\mathrm{sup}$.  The IMF is normalized such that
\begin{equation}
\int_{m_{inf}}^{m_{sup}}dm\ m\phi (m)=1\ ;
\end{equation}
(iv) the cosmic baryon accretion rate $a_\mathrm{b}(t)$, which
accounts for the increase in the fraction of baryons in structures
that have deep enough potential wells for  hydrogen to be  ionized,
from an initial fraction of $a_\mathrm{init}\sim$ 1 \% 
at $z=20$ in mini-halos \citep*[cf.][]{mo:02}
to the present value at $z=0$
in galaxies of $\sim 10-15 \%$ derived by \citet*{fukugita:98} and
\citet{dickinson:03}; (v) the gas outflow from the cosmic structures
$o(t)$.

The evolution of the gas mass in the IGM is given by\,:
\begin{equation}
\frac{d M_\mathrm{igm}}{dt} = -a_\mathrm{b}(t) + o(t)\ .
\end{equation}
We consider two possible parametrizations of the cosmic baryon
accretion rate $a_\mathrm{b}(t)$ \,: (i) very localized structure
formation at time $t_{0}$\,:
\begin{displaymath}
a_\mathrm{b}(t) = a M_\mathrm{tot}\ \delta(t-t_{0})\ ;
\end{displaymath} 
and (ii) an exponentially decreasing formation rate starting at
$t=0$\, :
\begin{displaymath}
 a_\mathrm{b}(t) = \frac{a}{\tau_\mathrm{s}}\ M_\mathrm{tot}\
\exp{\left(-t / \tau_\mathrm{s}\right)}\ ,
\end{displaymath}
where the total mass
$M_\mathrm{tot}=M_\mathrm{igm}+M_\mathrm{struct}$ is constant in the
simulation and the mass of the structures evolves as
\begin{displaymath}
\frac{d M_\mathrm{struct}}{dt} = a_\mathrm{b}(t) - o(t)\ .
\end{displaymath}
Initially $M_\mathrm{struct}/M_\mathrm{tot}=a_\mathrm{init}$ as
defined above.  In both cases, the constant $a$ is the fraction of the
total mass which is eventually accreted by the structures. It is
adjusted to have the correct final value of the baryon fraction in the
cosmic structures. In all of the models discussed below, we have
chosen $a = 10 $ \%.  In the second case, $\tau_\mathrm{s}$ is the
timescale of the accretion process by the structures : 95 \% of the
accretion process is finished at $t \sim 3\ \tau_\mathrm{s}$. 
As structure formation is not well constrained observationally 
 at high $z$, we consider a large sample of timescales 
 and show how they are constrained.
Below we
have considered four possible timescales : $\tau_\mathrm{s}=0.01$,
0.2, 0.5 and 1 Gyr.

The outflow contains two terms
$o(t)=o_\mathrm{w}(t)+o_\mathrm{sn}(t)$. The first one,
$o_\mathrm{w}(t)$, is a global outflow powered by the stellar
explosions (galactic wind). This term is similar to that described in
\citet{scully:97}. The mass ejection rate by the wind is given by
\begin{equation}
o_\mathrm{w}(t) = \epsilon \int_{\max{ \left(8\ \mathrm{M}_{\odot} ; m_\mathrm{d}(t)\right) }}^{m_\mathrm{up}}dm\ \phi(m) 
\psi\left(t-\tau(m)\right) \frac{2
E_\mathrm{kin}(m)}{v_\mathrm{esc}^{2}(m)}\ ,
\end{equation}
where $m_{up}$ is the upper mass limit of stars which expire as a
 supernova event and $m_\mathrm{d}(t)$ is the mass of a star which dies at age $t$. $E_\mathrm{kin}(m)$ and $v_\mathrm{esc}(m)$ are
 respectively the kinetic energy and the terminal velocity of the
 ejecta for a progenitor of mass $m$ obtained from \citet*{woosley:95,heger:02}, 
 and $\tau(m)$ is the main
 sequence lifetime of a star of mass $m$, taken here to be independent of metallicity. 
 The fraction of the energy
 available to drive the wind, $\epsilon = 0.02$, is taken to be constant. The
 second term, $o_\mathrm{sn}(t)$ corresponds to the fraction $\alpha$
 of stellar supernova ejecta which is flushed directly out of the
 structures, resulting in metal-enhanced winds as first proposed by
 \citet{vader:86}\,:
$$
o_\mathrm{sn}(t) = \alpha \int_{\max{ \left(8\ \mathrm{M}_{\odot} ; m_\mathrm{d}(t)\right) }}^{m_\mathrm{up}}dm\ \phi(m)
\psi\left(t-\tau(m)\right) \left(m-m_\mathrm{r}\right)\ ,
$$
where $m_\mathrm{r}$ is the mass of the leftover remnant and
the fraction $\alpha = 0.005$ is also taken to be constant.
Thus, as one can see, $o_\mathrm{w}(t)$ carries the chemical composition of the ISM, 
whereas $o_\mathrm{sn}(t)$ has  the chemical composition of the supernovae.

The evolution of the gas mass in the structures is given by\,:
\begin{equation}
\frac{d M_\mathrm{ism}}{dt} = \left( -\psi(t) + e(t)\right) +
\left(a_\mathrm{b}(t) - o(t)\right)\ .
\label{eq:dmismdt}
\end{equation}
The stellar mass in the structures is of course given by the
difference $M_\mathrm{struct}-M_\mathrm{ism}$.  In
equation~(\ref{eq:dmismdt}), the second parenthesis accounts for the
baryon exchange between the IGM and the structures, whereas the first
parenthesis is the classical term related to the star
formation. Specifically, the rate at which gas is returned to the ISM
by mass loss or stellar deaths either in supernova events or in
planetary nebulae, $e(t)$, is calculated including the effect of
stellar lifetimes (delaying enrichment, though these effects are
negligible for extremely massive stars), that is, we do not employ the
instantaneous recycling approximation. Therefore $e(t)$ is given
by\, :
\begin{equation}
e(t) = \int_{m_\mathrm{d}(t)}^{m_{up}}dm\ \phi (m)\psi (t-\tau (m))
(m-m_\mathrm{r})\ .
\label{eq:e}
\end{equation} 
Notice the distinction between
$m_\mathrm{sup}$, the upper mass limit of stars which can form, and
$m_\mathrm{up}$ which has been defined above.  It has been suggested
\citep*{larson:86,olive:87} that one way to avoid
the overproduction of $^{16}$O in chemical evolution models is to
limit the maximum mass of stars which will give a supernova.  Stars
more massive than this are assumed to collapse entirely into black
holes returning no material to the ISM. This idea is now supported by
the recent calculation of \citet*{heger:02} (see below).  As we
will see this distinction may be quite important for very massive
stars required for ionization, but may be problematic from the
point of view of chemical enrichment.

The chemical evolution of a specific element $i$ is also given by two
equations, one for the mass fraction in the IGM, $X_{i}^\mathrm{igm}$,
and one for the mass fraction in the ISM of the cosmic structures,
$X_{i}^\mathrm{ism}$\,:
\begin{equation}
\frac{d X_{i}^\mathrm{igm}}{dt} = \frac{1}{M_\mathrm{igm}(t)}\left[
 o_\mathrm{w}(t)
 \left(X_{i}^\mathrm{ism}(t)-X_{i}^\mathrm{igm}(t)\right) +\left(
 o_{\mathrm{sn}, i}(t)-o_\mathrm{sn}(t) X_{i}^\mathrm{igm}(t)\right) \right]\
 ,
\end{equation}
\begin{eqnarray}
\frac{d X_{i}^\mathrm{ism}}{dt} & = &
\frac{1}{M_\mathrm{ism}(t)}\left[ \left(e_{i}(t)-e(t)
X_{i}^\mathrm{ism}(t)\right)\right.\nonumber\\ & & \left.
+a_\mathrm{b}(t)\left(X_{i}^\mathrm{igm}(t)-X_{i}^\mathrm{ism}(t)\right)
-\left( o_{\mathrm{sn}, i}(t)-o_\mathrm{sn}(t) X_{i}^\mathrm{ism}(t)\right)
\right]\ ,
\end{eqnarray}
where $o_{\mathrm{sn}, i} / o_\mathrm{sn}$ is the mass fraction of
element $i$ in the supernova ejecta which are directly ejected in the
IGM and therefore do not contribute to the chemical enrichment of the
structures, and $e_{i}(t)$ is the rate at which element $i$ is ejected
by stars into the ISM. It is given by
\begin{equation}
e_{i}(t) = \int_{m_\mathrm{d}(t)}^{m_{up}}dm\ \phi (m)\psi (t-\tau
(m)) m^\mathrm{ej}_{i}(m).
\label{eq:ei}
\end{equation}
The ejected mass of element $i$ by a star of initial mass $m$,
 $m^\mathrm{ej}_{i}(m)$, is computed from the stellar yields. Despite
 the fact that this dependence has not been explicitly written in
 equations~(\ref{eq:e}) and (\ref{eq:ei}), the mass of the remnant
 $m_\mathrm{r}$, the ejected mass of element $i$, $m^\mathrm{ej}_{i}$,
 and for the massive stars the kinetic energy $E_\mathrm{kin}$ and the
 terminal velocity $v_\mathrm{esc}$ of the supernova ejecta are
 functions not only of the initial stellar mass $m$, but also of its
 initial metallicity, which is taken to be the metallicity of the ISM
 at time $t-\tau(m)$. Notice that for the specific case of iron, we
 add a term accounting for the production of radioactive nickel by
 type Ia supernovae, assuming that they typically occur with a delay of 1 Gyr and they
 produce 0.5 M$_{\odot}$ of nickel per event.

We have divided the stellar mass range in three domains : (i) for
intermediate mass stars ($< 8\ \mathrm{M}_{\odot}$), the remnant is
always a white dwarf. We adopt stellar lifetimes from \citet{maeder:89}. 
The stellar yields and the mass of the white dwarfs are
taken from \citet*{vandenhoek:97} for stars with masses in
the range 0.9 - 8 M$_\odot$ and metallicity in the range Z = 0.0001,
0.004, 0.008, 0.02; (ii) for massive stars ($8 < m < 100\
\mathrm{M}_{\odot}$), we adopt stellar lifetimes from 
\citet{schaerer:02}. We take the nature of the remnant (neutron star or black hole)
from \citet{heger:03} and the stellar yields as well as the mass of
the remnant from \citet*{woosley:95} for stars with mass in the
range $12-40\ \mathrm{M}_\odot$ and metallicity in the range
Z/Z$_\odot$ = 0, 0.0001, 0.01, 0.1, 1. Specifically we use the results
of their model B using a kinetic energy $E_\mathrm{kin}\sim 2\times
10^{51}\ \mathrm{erg}$ for $m=30$, $35$ and $40\
\mathrm{M}_{\odot}$, as this model leads to a better agreement with the
observed abundance ratios in extremely-metal poor halo stars (see section 4.3).
The maximum initial stellar mass studied in their
paper is $40\ \mathrm{M}_\odot$, which necessitates an extrapolation for
$40 < m < 100\ \mathrm{M}_\odot$.
We note that the slope of the IMF favors stars with lower masses, which means that the contribution of stars close to $100\ \mathrm{M}_\odot$ is always very small in all our models. For the same reason,
the arbitrary choice of $100\ \mathrm{M}_\odot$ as maximum mass is of little influence. Our results would be unchanged with a maximum mass of $140\ \mathrm{M}_\odot$ ;
(iii) finally for very massive
stars ($> 140\ \mathrm{M}_\odot$), we use the recent calculations of
\citet*{heger:02}, which predict that stars end as a so-called
PISN for $140 < m < 260\ \mathrm{M}_\odot$ and entirely collapse into
a black hole for $m > 260\ \mathrm{M}_{\odot}$. The corresponding
stellar yields are taken from the same paper. We adopt the stellar
lifetimes from \citet{schaerer:02}.  Note also that for $m<100\
\mathrm{M}_{\odot}$, the evolution of the metallicity dependence is
taken into account by an interpolation between the sets provided in
the studies cited above.\\

We investigate several scenarios of cosmic star formation\,:
\begin{enumerate}

 \item A Standard model (``Model 0'')
 
 The standard model which we use to compare the more speculative
 models involving very massive stars is defined by an IMF with a slope
 slightly steeper than that of a Salpeter mass function (x = 1.35)
\begin{equation}
\phi(m) \propto m^{-(x+1)} \qquad x = 1.7
\label{stimf}
\end{equation}
for stellar masses between 0.1 and 100 M$_\odot$ \citep{scalo:86}, and a
star formation rate (SFR) proportional to the gas mass fraction 
in the cosmic structures,
$\sigma(t)$,
\begin{equation}
\psi(t) = \nu_{1} \sigma(t)\ ,
\end{equation}
where $\nu_{1} = M_\mathrm{struct}(t) / \tau_{1} $ with $\tau_{1} = 5\
\mathrm{Gyr}$, which is the typical timescale for star formation in
the galactic disk.
The overall (late) evolution of the chemical abundances are quite
acceptable in this model.  However, as we shall see, this model can
not explain the rather peculiar nature of the extreme metal poor star
HE 0107-5240 nor can this model achieve the necessary degree of
ionization at high redshift.

 \item A massive mode added to a standard-like model (``Model 1'')
 
 In this case, we consider an epoch of massive star formation with a
 power-law IMF as in eq. (\ref{stimf}) but with a range 40 - 100
 M$_\odot$.  The brevity of this period of star formation is governed
 by the SFR which is taken here to be an exponential,
 \begin{equation}
\psi(t) = \nu_{2}\ e^{-t/\tau_{2}},
\label{mmsfr}
\end{equation}
where $\tau_{2}=50\ \mathrm{Myr}$ is the characteristic timescale of
this massive starburst and $\nu_{2} = f_{2}\ M_\mathrm{struct}(t) / \tau_{2}$. The fraction $f_{2}$ equals 4.5 \% 
so that $\nu_{2} = \left(0.9\ \mathrm{Gyr}^{-1}\right) M_\mathrm{struct}(t)$.  This short
period of star formation activity acts as a prompt initial enrichment
of the cosmic structures which leads rapidly to a metallicity as high
as $10^{-3} \ Z_{\odot}$ and at the same time provides ample
energy for the ionization of the IGM. In this context, the possible
delay between the massive and normal SFR is required to be very short
(of the order of the lifetime of a typical massive star, a few $\times
10^{6}$ yrs).  So there is no significant difference between the
simultaneous combination of the two modes or the introduction of this
short delay which affects any correlation between nucleosynthesis and
the rate of ionization.  This delay can
account for a period of negative feedback due to the lack of neutral
H$_2$ necessary for cooling \citep{bromm:04}.  The chemical history of the
elements is in good agreement with observables between $z=5$ to $z=0$,
including the essential of observations of HE017-5240.

\item Pair Instability SN mode (``Model 2a'')

 We also consider a more extreme case, where the initial massive mode
(once again given by the same power-law IMF) covers stellar masses in
range 140 - 260 M$_\odot$, corresponding to the so-called Pair
Instability supernovae.  We take the same SFR as in eq. (\ref{mmsfr})
and the yields are taken from \citet{heger:02}.

In this model, for the same SFR
 intensity, we find a larger production of metals  
 than we did in case 2 due to the fact no remnant is left
 behind and all of the enriched stellar mass is returned to the ISM
 and IGM.  In this context, we must lower the SFR of the massive mode
 by a factor of about 8 in order to avoid the overproduction of
 metals in the ISM.  Consequently, the ionizing efficiency is lower than in case
 2. Finally, due to the specific nucleosynthesis ascribed to PISNae,
 we can not reproduce the abundance patterns (e.g., of C, N, Zn )
 found in the data of the HE 0107-5240.

\item Black hole mode (``Model 2b'')

 Finally we consider a mode where the massive IMF covers stellar
 masses in the 270 - 500 M$_\odot$ range. All of these stars are
 assumed to collapse entirely to black holes without contaminating the
 environment.  Indeed, the ill effects on the production of heavy
 elements by the extreme massive component of the case 3 can be
 ameliorated if it is assumed that these stars provide only a source
 of ionization.  In this case, there is a de-correlation between
 global chemical evolution and reionization.  Chemical enrichment
 could proceed solely through the normal component. Although this mode
 fit observed abundances well at late times, it does not fit the
 abundances which are characterized by HE 0107 5240 at low [Fe/H].  We
 further note that the SFR of this very massive mode is in principle constrained from above, and hence we can not adjust it arbitrarily to obtain reionization. Indeed, if a large fraction of
the baryons in the structures was trapped in black holes, the remaining reservoir
to form stars would not be sufficient to fit the observed cosmic SFR.
Moreover, the intensity of this early burst of very massive stars,
 governed by the fraction $f_{2}$, 
must also be limited so that the contribution to the cosmic SFR
at $z\simeq 5-6$ of the tail of this massive mode remains negligible.
\end{enumerate}

\section{Reionization}

The ionizing flux from the stars is entirely dominated by the
contribution of massive stars with short lifetimes. Therefore, it is a
good approximation to consider that the total number of UV photons
emitted by a star of initial mass $m$, $N_{\gamma}(m)$, is entirely
emitted impulsively when the stars dies. Then the ionizing flux is
given by
\begin{equation}
\frac{dn_{\gamma}}{dt} = \int_{m_\mathrm{d}(t)}^{m_\mathrm{sup}}dm\
\phi(m) \psi\left(t-\tau(m)\right) N_{\gamma}(m)\ .
\end{equation}
We do not distinguish the reionization of hydrogen and helium in this
paper, and hence, we consider all UV photons with $h\nu \ge 13.6\
\mathrm{eV}$. We adopt the values of $N_{\gamma}(m)$ for stars with masses in
the range $5$--$500$ M$_{\odot}$ computed by \citet{schaerer:02} at zero metallicity. 
Only a fraction $f_\mathrm{esc}$ of these ionizing photons will escape the
structures and contribute to the reionization of the IGM.  To a first
approximation, we consider in this paper that $f_\mathrm{esc}$ is
constant, for lack of any better model.
Estimates of   $f_\mathrm{esc}$ range from $\sim 0.03$ to  $\sim 0.5,$
the former applying to nearby star-forming galaxies and the
latter applying to Lyman break galaxies at $z\sim 3$ 
\citep*{steidel:01}. Note that we do not include any 
photon-ionization feedback on the star formation 
as this effect is expected to be weak at very high
redshifts (\citealt*{oh:03}; \citealt{dijkstra:04a}).
 
In a forthcoming
paper, we will perform the detailed calculation of the reionization process
in the IGM, taking into account the balance between the ionizing flux
and the recombination rate for each species (H$^{+}$, He$^{+}$ and
He$^{++}$). 
In this paper we adopt a simplified criterion to decide
whether our stellar population is able to reionize the early Universe or not.
We estimate the mean recombination time $\left\langle t_\mathrm{rec} \right\rangle$ of an atom ionized at redshift $z$ using 
\citep*{ricotti:04b}
\begin{equation}
\frac{\left\langle t_\mathrm{rec}\right\rangle}{t_\mathrm{H}} = 
0.046\ \left(\frac{1+z}{18}\right)^{-1.5}\ \left(\frac{C_\mathrm{H\;\scriptscriptstyle{II}}}{10}\right)^{-1}\ ,
\label{eq:trec}
\end{equation}
where $t_\mathrm{H}$ is the Hubble time at redshift $z$ and $C_\mathrm{H\;\scriptscriptstyle{II}}$ is the clumpiness factor of the ionized regions, whose evolution with redshift remains quite uncertain. We can estimate the number of ionizing photons per intergalactic baryon necessary to fully reionize the IGM at redshift $z$ with
\begin{eqnarray}
n_{\gamma,\ \mathrm{min}} & = & 
\left(\frac{\left\langle t_\mathrm{rec}\right\rangle}{t_\mathrm{H}}\right)^{-1} = 22\ \left(\frac{1+z}{18}\right)^{1.5}\ \left(\frac{C_\mathrm{H\;\scriptscriptstyle{II}}}{10}\right)\ \mathrm{ph/b}\  \mathrm{if}\ \left\langle t_\mathrm{rec}\right\rangle < t_\mathrm{H}\nonumber\\
& = & 1\ \mathrm{ph/b}\ \mathrm{otherwise}
\label{eq:nrec}
\end{eqnarray}
As such, we will consider that
early reionization at $z\sim 17$ is fully achieved only
if the structures have cumulatively 
ejected $\sim 10-20$ ionizing photons per intergalactic baryon by the end of the initial
massive starburst (i.e. typically for $z_\mathrm{init}=20$).
While the reionization redshift increases with the number of ionizations per
photon, it is reduced by the mean clumpiness of the IGM.
If the  first generation of stars formed in isolated star cluster-mass
structures, which
seems plausible at high redshift in a hierarchical model of structure
formation, $f_\mathrm{esc}$ is likely to be high \citep{ricotti:02}.
We  therefore conservatively take as the required production rate of ionizing
photons per baryon a number of about 10. See for example a recent
discussion \citep*{dijkstra:04b}
that estimates  $\sim 10$ ionizations per intergalactic baryon are needed
to maintain a constant ionization fraction for a specified  optical
depth taken equal to the WMAP estimate and a clumpiness factor of
 the IGM
$\langle n^2\rangle/\langle n \rangle ^2 \sim 3$.

\section{Results}

In this section we display and compare the results of the models
considered.  We will discuss in turn,  the star
formation rate, the ionization flux and the chemical evolution of the
element abundances.  In all of the models considered, star formation
begins at $z_\mathrm{init} = 20$.

\subsection{The Star Formation Rate}

We begin by first comparing the results for each of the models with
respect to their predicted star formation history.  As discussed above
we will consider several scenarios for the time evolution of the
baryon accretion rate. In Model 0, we both adopt a uniform baryon
accretion rate between $z = 20$ and $z = 0$ and an exponentially
decreasing baryon accretion rate with several timescales
$\tau_\mathrm{s}$ : 0.01, 0.2, 0.5 and 1 Gyr. For models 1 and 2, we
consider only the latter, however in the case of Model 1, we also show
the effect of a delta function at $z=6$ (late and rapid structure
formation).

\begin{figure}
\begin{center}
\plotone{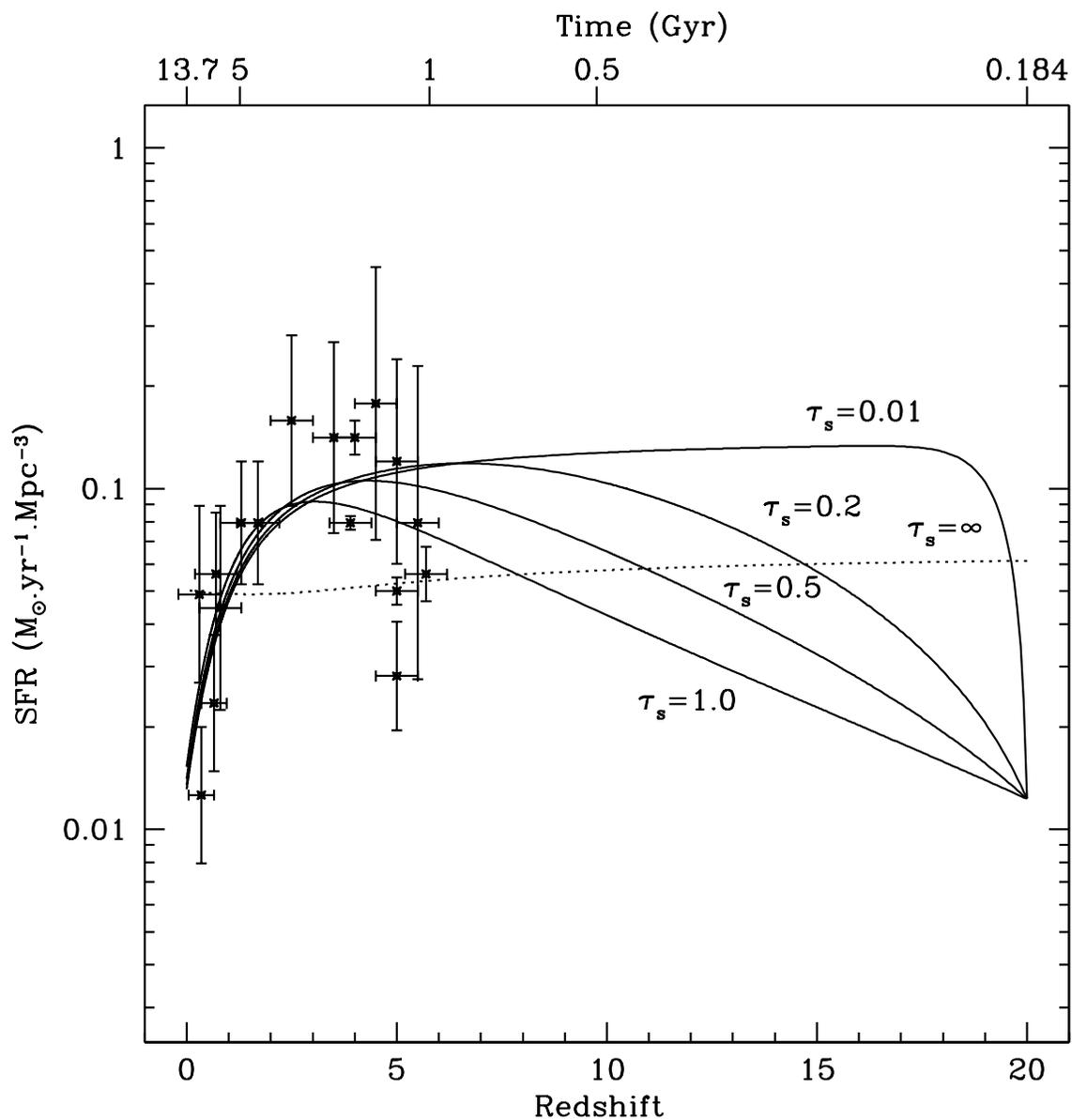}
\end{center}
\caption{\textbf{Cosmic star formation rate : standard model (Model
0).} The observed cosmic SFR is taken from \citet{lilly:96},
\citet{pascarelle:98}, \citet{iwata:03}, \citet{ouchi:03} and
\citet{giavalisco:04}. We consider two possible scenarios for
structure formation : either a uniform baryon accretion rate from
$z=20$ to $z=0$ (dotted line) or an exponentially decreasing accretion
rate (solid line) with a timescale $\tau_\mathrm{s}$ as labelled.}
\label{fig:sfr0}
\end{figure}

 In Figure~\ref{fig:sfr0} we show the evolution of the cosmic star
formation rate for Model 0, and compare it to recent observations
(deduced from luminosity functions) \citep[][and references therein]{nagamine:03}.  
We plot the comoving star formation rate in
units of M$_\odot$yr$^{-1}$Mpc$^{-3}$ which is related to $\psi(t)$
by
\begin{displaymath}
\mathrm{SFR}\ \mathrm{(M_{\odot}yr^{-1}Mpc^{-3})} =
\rho_\mathrm{b}\ \mathrm{(M_{\odot}Mpc^{-3})}\ \times\
\frac{\Psi(t)}{M_\mathrm{tot}}\ \mathrm{(yr^{-1})},
\end{displaymath}
where $\rho_\mathrm{b} = 2.77\times 10^{11}\
\left(\Omega_\mathrm{b} h^{2}\right)\
\mathrm{M}_{\odot}\mathrm{Mpc}^{-3}$ is the comoving baryon density
in the Universe. We take $\Omega_{b}=0.044$.
 With the exception of the constant accretion rate ($\tau_\mathrm{s} =
 \infty$, shown as the dotted curve) all of the models with
 $\tau_\mathrm{s} = 0.01 - 1.0$ Gyr reproduce the observed rise
 of the SFR between $z=0$ and $z=5$.  Note that in the case of a 
 constant accretion rate  ($\tau_\mathrm{s}=\infty$), slower
 structure formation necessitates an intrinsically more efficient rate of star
 formation in order to correctly reproduce the global chemical
 evolution history. Therefore we have multiplied the initial SFR intensity by a factor
 of $\sim 5$ so that $\tau_{1}=1$ Gyr for this case. This explains why in
 Figure~\ref{fig:sfr0} the initial value of the dotted curve is $\sim
 5$ times larger than it is  for the other models.

Despite  the large observational uncertainty, the shape of the observed
SFR up to $z=5$ allows us to constrain the completion time scale for the formation of
structures, i.e.  $\tau_\mathrm{s} \la 1$ Gyr.
When this constraint is satisfied, Model 0 is in good agreement with
the observed behaviour of the SFR at late times ($z<5$). On the other
hand, the lack of data at high redshift does not allow one to
distinguish between rapid  ($\tau_\mathrm{s}=0.01\
\mathrm{Gyr}$) and prolonged ($\tau_\mathrm{s}\sim 0.2-0.5\
\mathrm{Gyr}$) structure formation.  The recent discovery of a candidate $z=10$ galaxy
\citep{pello:04,ricotti:04a} indicates that SFR determination at $z\le 10$
should become available soon.

\begin{figure}
\begin{center}
\plotone{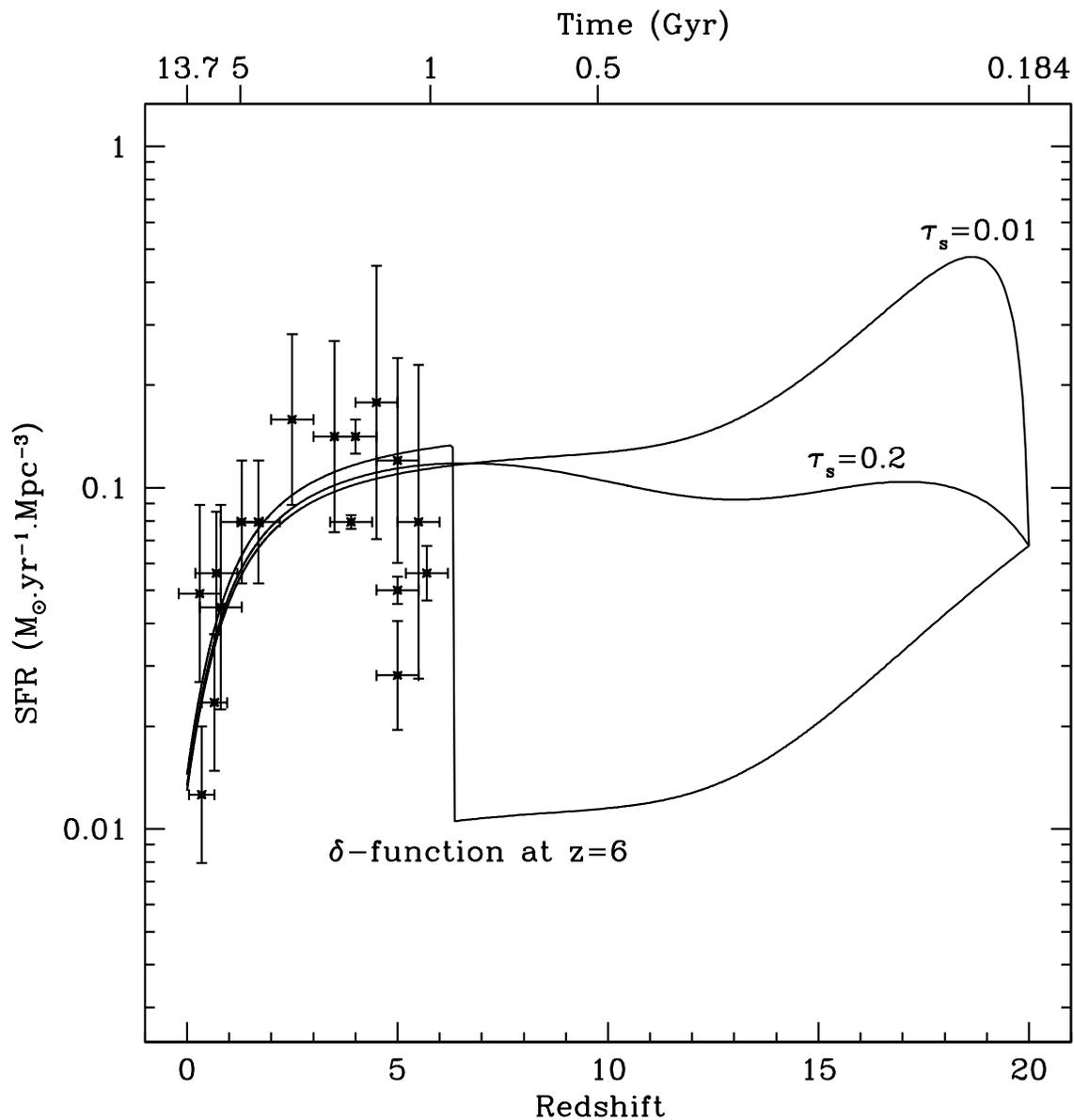}
\end{center}
\caption{\textbf{Cosmic star formation rate : standard model + massive
mode (Model 1).} Same as in figure~\ref{fig:sfr0}. The two envisaged
scenarios for structure formation are now either an exponentially
decreasing accretion rate with timescale $\tau_\mathrm{s}$ or a late
impulse of structure formation at $z=6$.}
\label{fig:sfr1}
\end{figure}

 The bimodal SFR defining Model 1 is a linear combination of the
 normal mode used in Model 0 and a second mode, favouring massive
 stars at the onset of star formation.  This massive mode is
 characterized by a SFR which is not coupled to the gas fraction in
 the structures but is simply exponentially decreasing with a timescale
 $\tau_{2}=0.05\ \mathrm{Gyr}$ and a lower mass limit of the IMF,
 $m_\mathrm{inf}=40$ M$_{\odot}$. Figure~\ref{fig:sfr1} shows the
 cosmic SFR obtained in Model 1 for three possible histories of 
 structure formation.  As in Model 0, we consider an exponentially
 decreasing baryon accretion rate with timescales $\tau_\mathrm{s}$ =
 0.01 or 0.2 Gyr.  Compared to Figure~\ref{fig:sfr0}, where the
 initial value of the SFR is $1.2\times 10^{-2}\
 \mathrm{M}_{\odot}\mathrm{yr}^{-1}\mathrm{Mpc}^{-3}$, the initial
 value of the SFR in Figure~\ref{fig:sfr1} equals $6.7\times 10^{-2}\
 \mathrm{M}_{\odot}\mathrm{yr}^{-1}\mathrm{Mpc}^{-3}$, which
 corresponds to an increase by a factor $(\nu_{2}+\nu_{1})/\nu_{1}=
 5.5$ due to the contribution of the massive mode.
We also consider a model in which the accretion occurs instantaneously to model 
a burst of star formation
relatively late at $z=6$ (this model is
quoted as ``delta-function at $z=6$'' in Figure~\ref{fig:sfr1}). As we
see, the late time evolution of the models is dominated by the normal
mode. For this reason, as in Model 0, the observed SFR is well
reproduced in all of the scenarios for $z\la 5$. As noted above for
Model 0, it is impossible to distinguish between the three scenarios
on the basis of SFR observations alone. However we will see in
section~\ref{sec:photons} that the early ionizing flux in the case of instantaneous late
structure formation is not able to ionize the IGM.

Models 2a and 2b have been considered in order to study the case of an
initial starburst with a more extreme mass range. Model 2a corresponds
to a starburst forming massive stars between 140 and 260 M$_{\odot}$,
which is the domain where stars end their life as PISNae -- the star
is entirely destroyed leaving no remnant and a very large amount of
metals is released into the ISM. In contrast, Model 2b corresponds to
a even more massive domain, $270 \le m \le 500\ \mathrm{M}_{\odot}$
where stars collapse entirely into black holes, without any
contribution to the metal enrichment of the surrounding gas. The SFR
in both cases is taken to be the same as in Model 1, with a timescale
$\tau_{2}=0.05$ Gyr. In this case, we only consider  the fast structure
formation mode ($\tau_\mathrm{s}=0.01$ Gyr) as it optimizes the early
emission of ionizing photons.

\begin{figure}
\begin{center}
\plotone{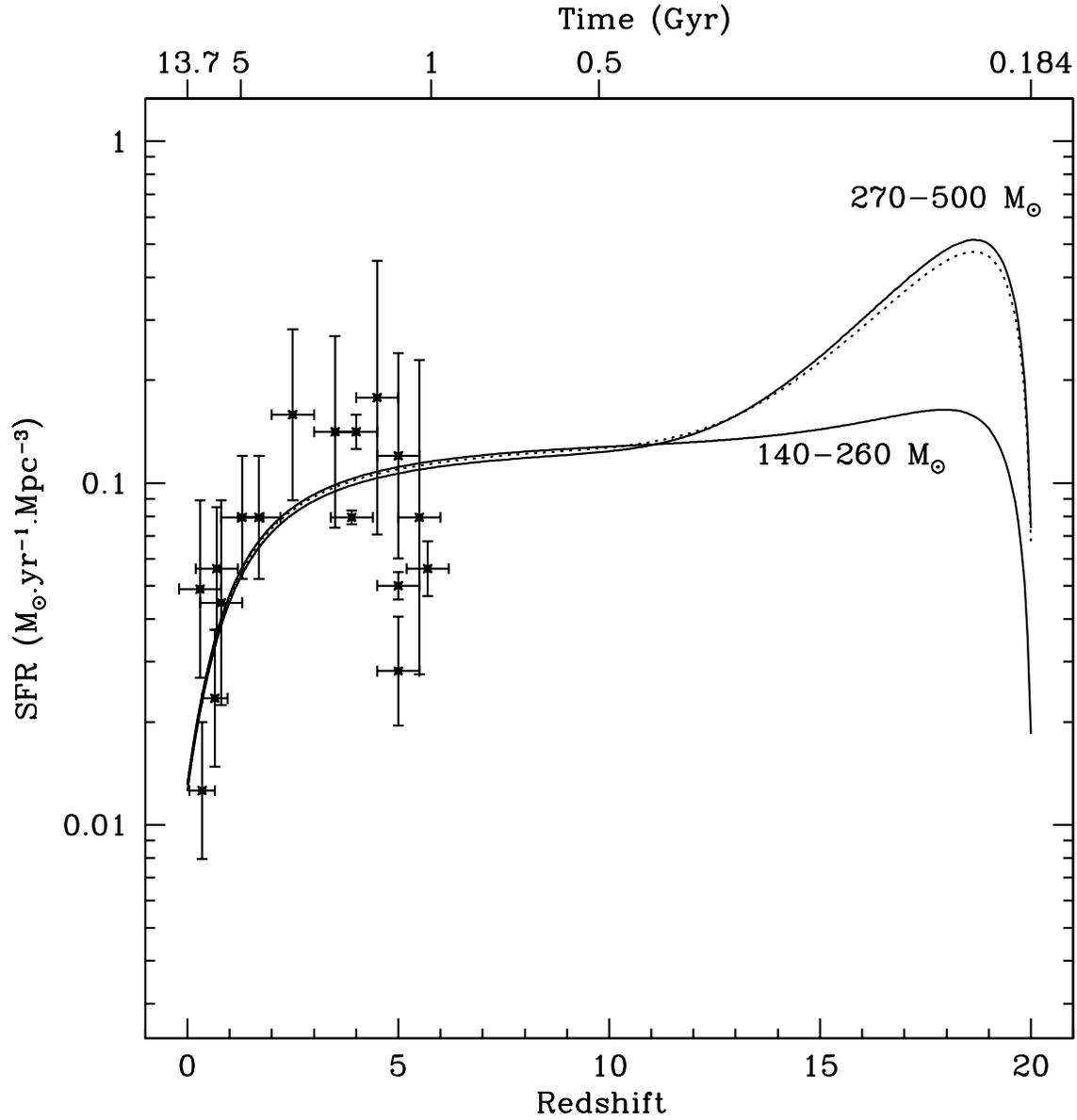}
\end{center}
\caption{\textbf{Cosmic star formation rate : standard model + very
massive mode (Models 2a and 2b).} Same as in
figure~\ref{fig:sfr0}. Models 2a and 2b are labeled with their mass
range. Model 1 with $\tau_\mathrm{s}=0.01$ has been plotted (dotted
line) for comparison.}
\label{fig:sfr2}
\end{figure}

The two mass domains lead to very different results. In Model 2a,
because of the intense production of metals by the PISN stars, it is
necessary to decrease the intensity of the massive starburst by a
factor of about 8 to avoid overabundances. Therefore we take
$f_{2}=0.56$ \% 
corresponding to $\nu_2 = \left(0.11 \ \mathrm{Gyr}^{-1}\right) M_\mathrm{struct}(t)$. 
As the late evolution
is dominated by the normal mode, the computed cosmic SFR is still in
good agreement with the observations, as can be seen in
Figure~\ref{fig:sfr2}.  For Model 2b, Figure~\ref{fig:sfr2}
corresponds to the maximum allowed starburst intensity which still
allows the normal mode to dominate below $z=5$. 

\subsection{The Ionization Flux}
\label{sec:photons}

\begin{figure}
\begin{center}
\plotone{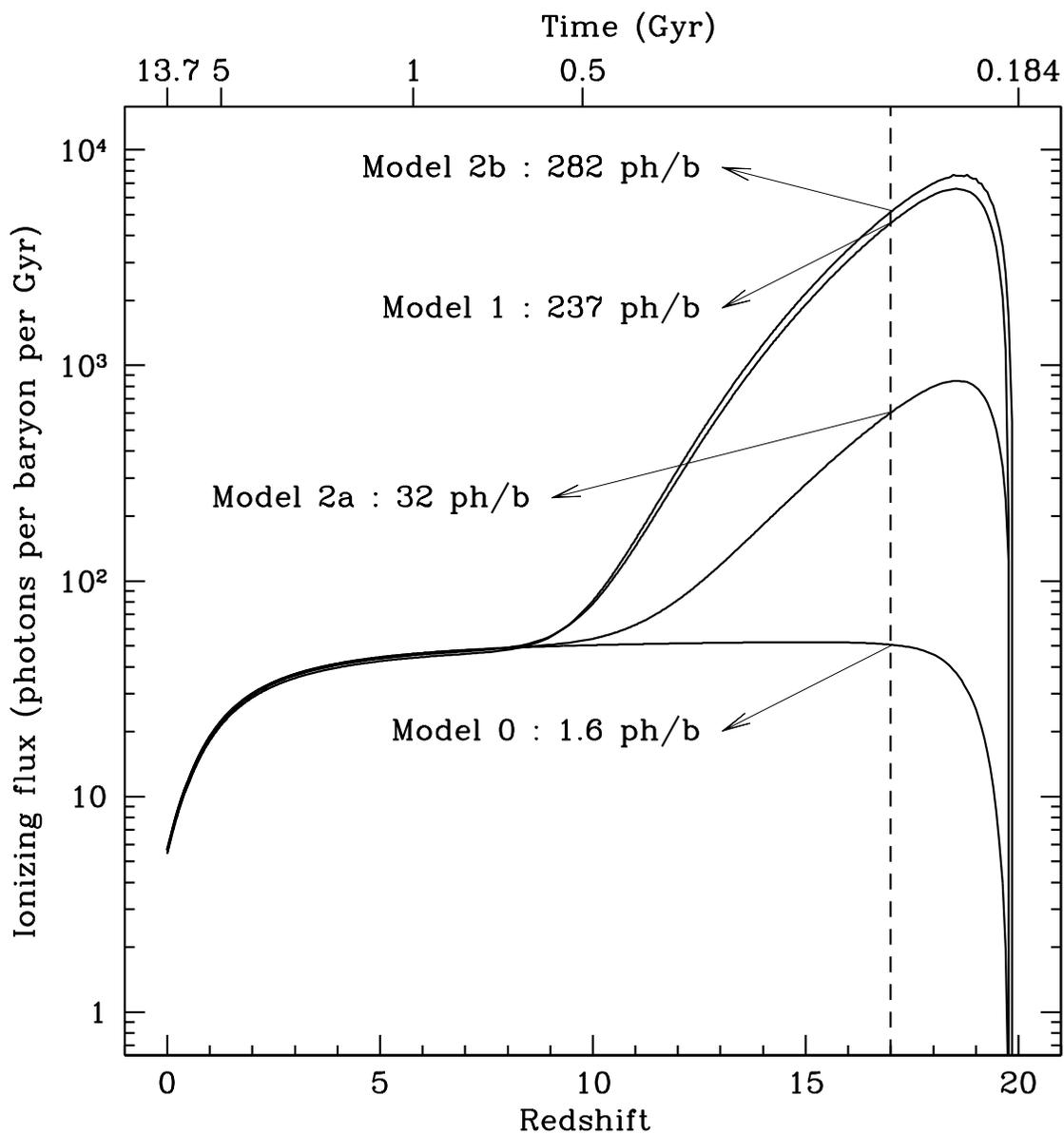}
\end{center}
\caption{\textbf{Reionization.} The ionizing flux is plotted as
function of redshift for the four models considered (0,
1, 2a and 2b). The total number of ionizing photons per intergalactic baryon
produced at $z=17$ is labeled for each curve. Only a fraction
$f_\mathrm{esc}$ of these photons are available for the early
reionization of the Universe, which is possible above $\sim$ 10
photons per intergalactic baryon (see Figure~\ref{fig:photons}).}
\label{fig:flux}
\end{figure}

The structure formation timescale has a direct impact on the early
ionizing flux of photons, as it governs the size of the gas reservoir
for star formation. Therefore we restrict our attention to
$\tau_\mathrm{s}=0.01$ Gyr for Model 0, which optimizes this
flux. However, even in this case, the number of ionizing photons
produced at $z=17$ (1.6 photons per baryon) is much too low to
reionize the early Universe (see Figure~\ref{fig:flux}). For this
reason, we conclude that Model 0 can be rejected and that a massive
mode is required.

Indeed, we see in Figure~\ref{fig:flux} the results for the ionizing
photon flux for models 1 and 2 (a and b).  The early ionizing flux is
strongly dependent on the early structure formation history. In Model
1, our calculations show that it is only when structure formation is
rapid (exponential decay with $\tau_\mathrm{s}=0.01$ Gyr) that
early reionization of the Universe is possible. Figure~\ref{fig:flux}
shows indeed that in this case, 237 photons per baryon are produced at
$z=17$, which allows for reionization with a low value of $f_\mathrm{esc}
\simeq 5\ \%$. In contrast, the case $\tau_\mathrm{s}=0.2$ Gyr, which
is not plotted, produces an integrated ionizing photon flux of only  55
photons per baryon. Even worse, the case of the late structure
formation at $z=6$ plotted in Figure~\ref{fig:sfr1} provides only 
30 photons per baryon. Models with slow structure formation
would then require a high value of $f_\mathrm{esc}$ (20-60 \%).

In the case of Model 2a, the decrease of the massive starburst
intensity discussed above has a large impact on the number of ionizing
photons produced at $z=17$. As seen in Figure~\ref{fig:flux}, the flux
is reduced to only 32 photons per baryon. This makes the early
reionization of the Universe possible only if the fraction of photons
that escape the structure is at least as high as $1/3$. Model 2b, on
the other hand, yields results which are very similar to those of
Model 1. The resulting ionizing photon flux is, as expected, quite
high : 282 photons per baryon at $z=17$ which allows for the early
reionization of the Universe even with an escape fraction of  only
$f_\mathrm{esc}\sim 3.5 \%$.
The similarity between the early production of ionizing photons 
in Models 1 and 2b is related to the constraint imposed on intensity of the 
initial starburst in Model 2b by the SFR 
(see Figure~\ref{fig:sfr2}).

\begin{figure}
\begin{center}
\plotone{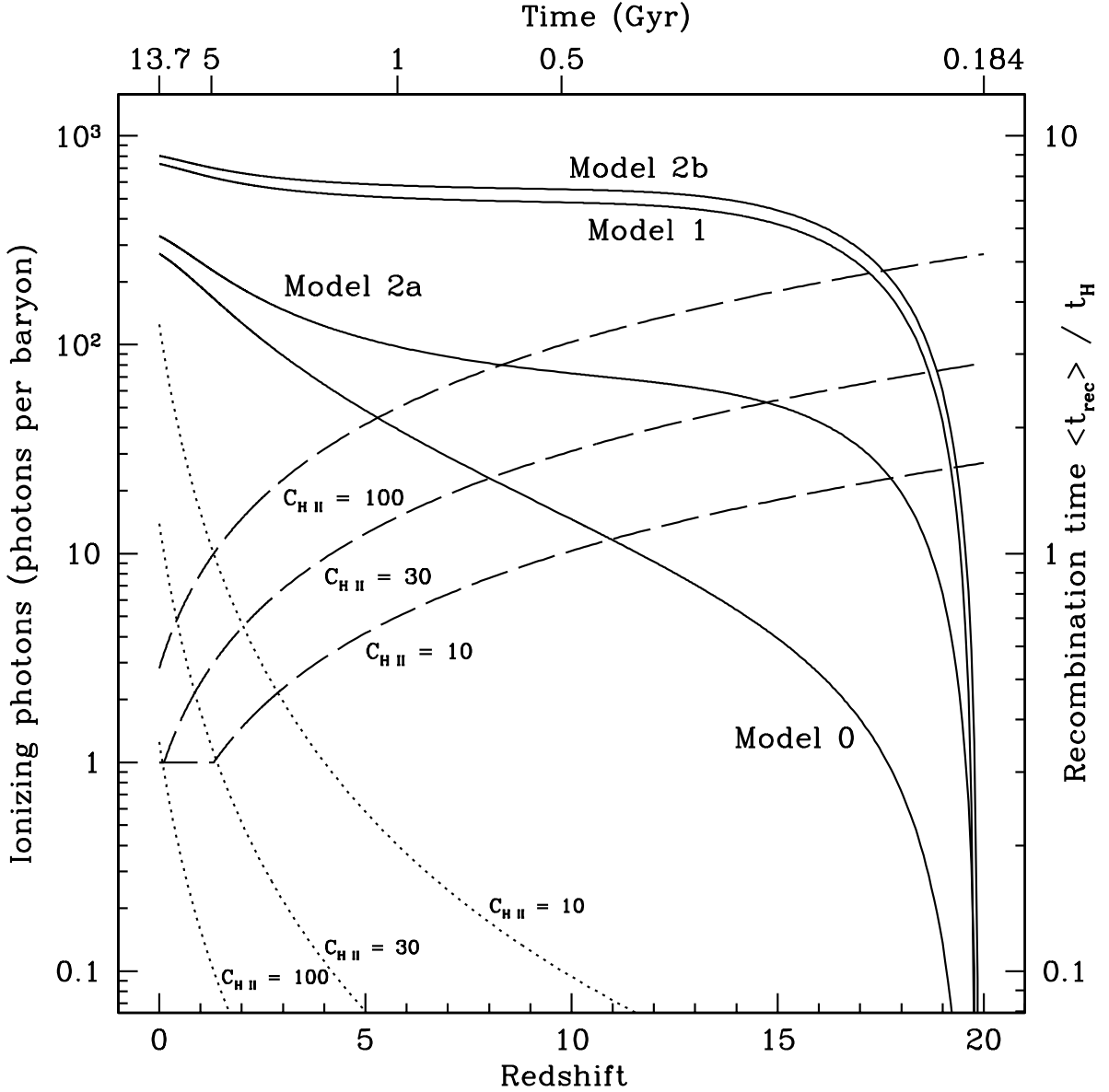}
\end{center}
\caption{\textbf{Reionization.} The total number of ionizing photons per intergalactic baryon is plotted as function of redshift for the four models considered (solid lines : 0, 1, 2a and 2b).
 Only a fraction $f_\mathrm{esc}$ of these photons are available to reionize the IGM.
The full reionization of the IGM is possible at redshift $z$ if the number of UV photons per intergalactic baryon produced by stars has reached a minimum value which is plotted by the dashed lines. The minimum number is computed from the mean recombination time of an atom ionized at redshift $z$, and  is plotted by the dotted lines in units of the Hubble time.
 Three values of the clumpiness factor of the ionized regions have been considered and are labeled in the figure : $C_\mathrm{H\;\scriptscriptstyle{II}}=10$, 30 and 100. }
\label{fig:photons}
\end{figure}

The capacity of Model 1 and Model 2b to reionize the IGM is confirmed in 
Figure~\ref{fig:photons}, where we have plotted
the cumulative number of ionizing photons as a function of redshift. 
We have also indicated the mean recombination time and the mean 
number of ionizing photons per baryon necessary to reionize the 
Universe at redshift $z$, as estimated using equations~(\ref{eq:trec}) 
and (\ref{eq:nrec}). The latter depends on the value of the clumpiness factor 
in \ion{H}{2} regions (which has been assumed to be constant with $z$ 
in this simple calculation). Model 2a marginally appears to be  
able to  reionize the IGM early. However, this model suffers from an 
additional difficulty : the peak of the ionizing flux (see Figure~\ref{fig:flux}) 
is reached at $z\sim 18-19$ where the metallicity is already  as 
high as $Z/ Z_\mathrm{\odot} \sim 10^{-2}-10^{-1}$. The possibility of  
forming stars with $m>140\ \mathrm{M}_{\odot}$ at these high metallicities 
is questionable \citep*[see e.g.][]{schneider:02,bromm:01b,ricotti:04b}. Although the 
metallicity in Model 1 is similar when the ionizing flux is maximum, it does not suffer 
this inconsistency as the IMF is truncated at $m=100\ \mathrm{M}_{\odot}$, 
and star formation is not as inhibited when $Z$ increases.
On the other hand, the very massive IMF 
($m>270\ \mathrm{M}_{\odot}$) in Model 2 does not produce 
metals (see next section). Therefore the metallicity is 
only $Z/Z_{\odot}\sim 10^{-4}-10^{-3}$ when the ionizing 
flux is maximum, i.e. consistent with  a very massive IMF.

As discussed in the next section, constraints from chemical evolution require that the 
massive starburst finishes by $z\sim 8-10$. 
After which, the normal mode is dominant and is less efficient at ionizing the IGM
(see Figure~\ref{fig:flux}). However the
 ionization of the IGM at these redshifts can be maintained 
despite the reduced number of photons per baryon. Thus,
it is not unlikely that a normal stellar population may account for the ionization 
level at $z \sim 6$, when the neutral gas fraction is $\ga 0.001$ \% 
\citep{oh:02},
as inferred from the observation of the Gunn-Peterson effect in the spectrum of high-redshift quasars. This possibility will be studied in a forthcoming paper presenting a detailed calculation of the reionization process.

\subsection{Element Abundances}

\begin{figure}
\begin{center}
\plotone{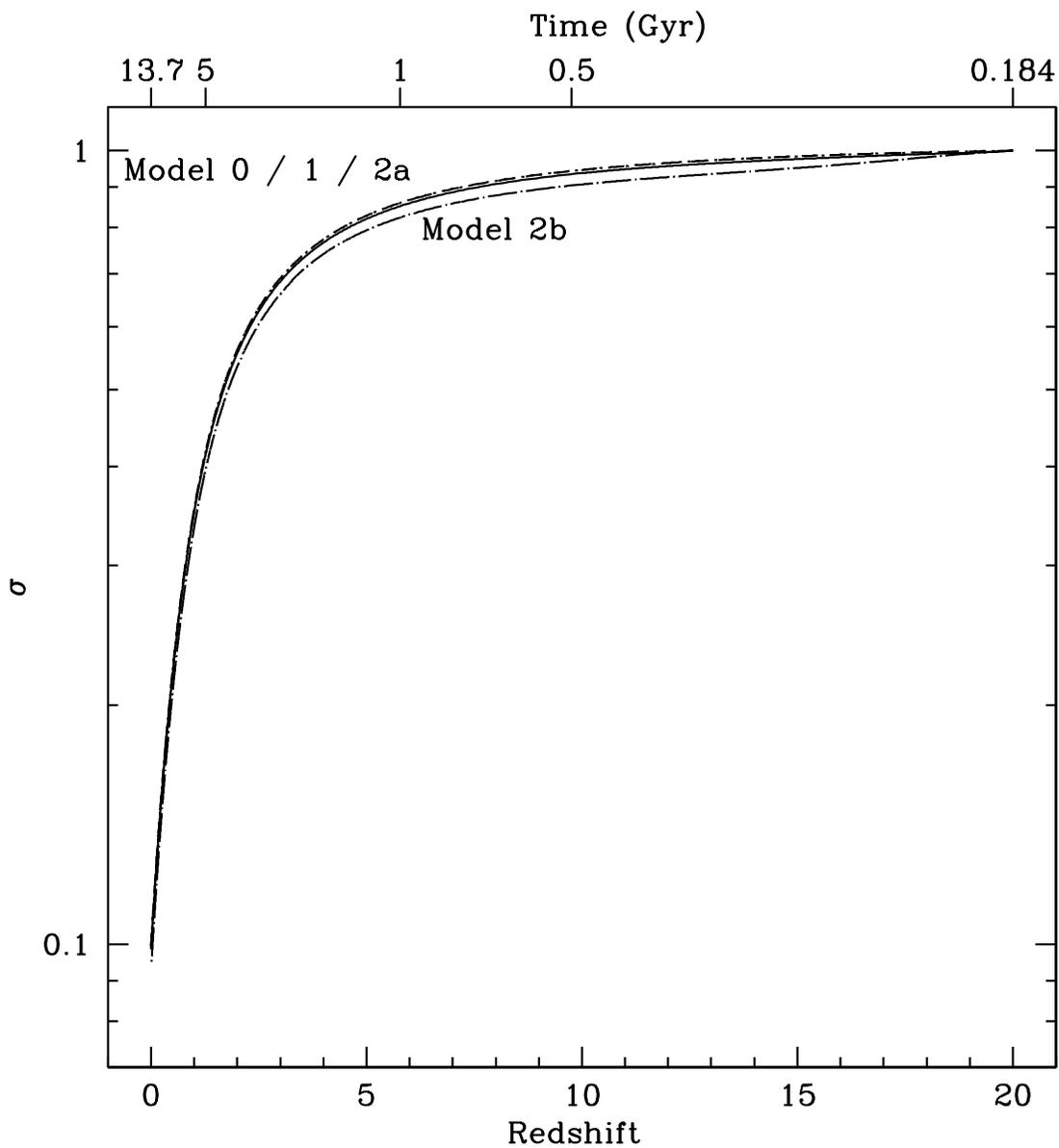}
\end{center}
\caption{\textbf{Evolution of the gas fraction in the cosmic
structures.} The gas mass fraction, $\sigma$, is plotted as a function of
redshift for Model 0 (dotted line), Model 1 (solid line), Model 2a
(dashed line) and Model 2b (dot-dashed line). Note that Models 0 and
2a are indistinguishable on this plot.}
\label{fig:sigma}
\end{figure}

\begin{figure}
\begin{center}
\plotone{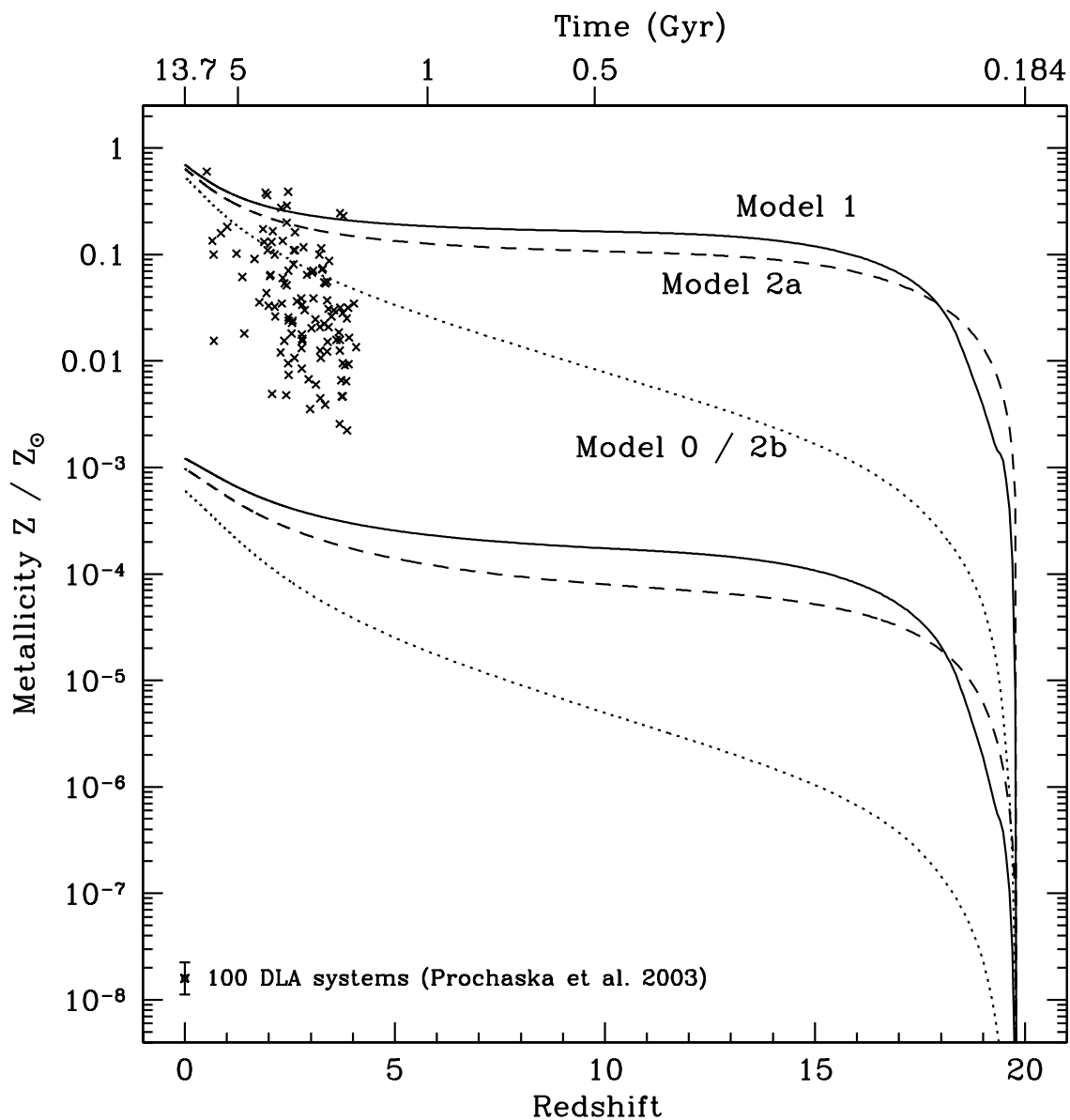}
\end{center}
\caption{\textbf{Evolution of the global metallicity.} The metallicity
in units of the solar metallicity is plotted as a function of redshift
for three models : Model 0 (dotted line), Model 1 (solid line) and
Model 2a (dashed line) both in the ISM of the cosmic structures (upper
curves) and in the IGM (lower curves). The predictions of Model 2b are
exactly the same as in Model 0.}
\label{fig:Z}
\end{figure}

Next, we turn our attention to the resulting element abundances in the
models considered.  We note first that the evolution of the gas
fraction in each of the models is quite similar as shown in
Fig.~\ref{fig:sigma}. In contrast, the total metallicity is quite
different in each case as seen in Fig.~\ref{fig:Z}. In
Fig.~\ref{fig:Z}, we show the evolution of the total metallicity
relative to solar metallicity in both the ISM of the growing
structures and the IGM. We also show the data from 100 DLAs at $z\le
4$ \citep{prochaska:03} which can be compared with the computed ISM
metal abundance.  As one might expect, Model 0 (and by definition
Model 2b) produce very reasonable results.  However, as we have just
shown that the addition of an initial massive starburst is necessary
for the early reionization of the Universe, it is now important to
check that the cosmic chemical evolution is still consistent both with
the local\footnote{ 'Local abundance'  refers to the mean abundance at $z=0$.} and high redshift observed abundances.
Figure~\ref{fig:Z} also shows that the metal enrichment is rapid in
 Model 1\,: $Z=10^{-3}-10^{-2}\ Z_{\odot}$ immediately at $z=20-19$
 \citep[see also][]{bromm:04}. As mentioned in the previous section, this
 justifies that the delay between the starting epochs of the two modes
 of star formation can be neglected in this context.  The global
 metallicity in Model 2a is very similar to that in Model 1.

\begin{figure}
\begin{center}
\plotone{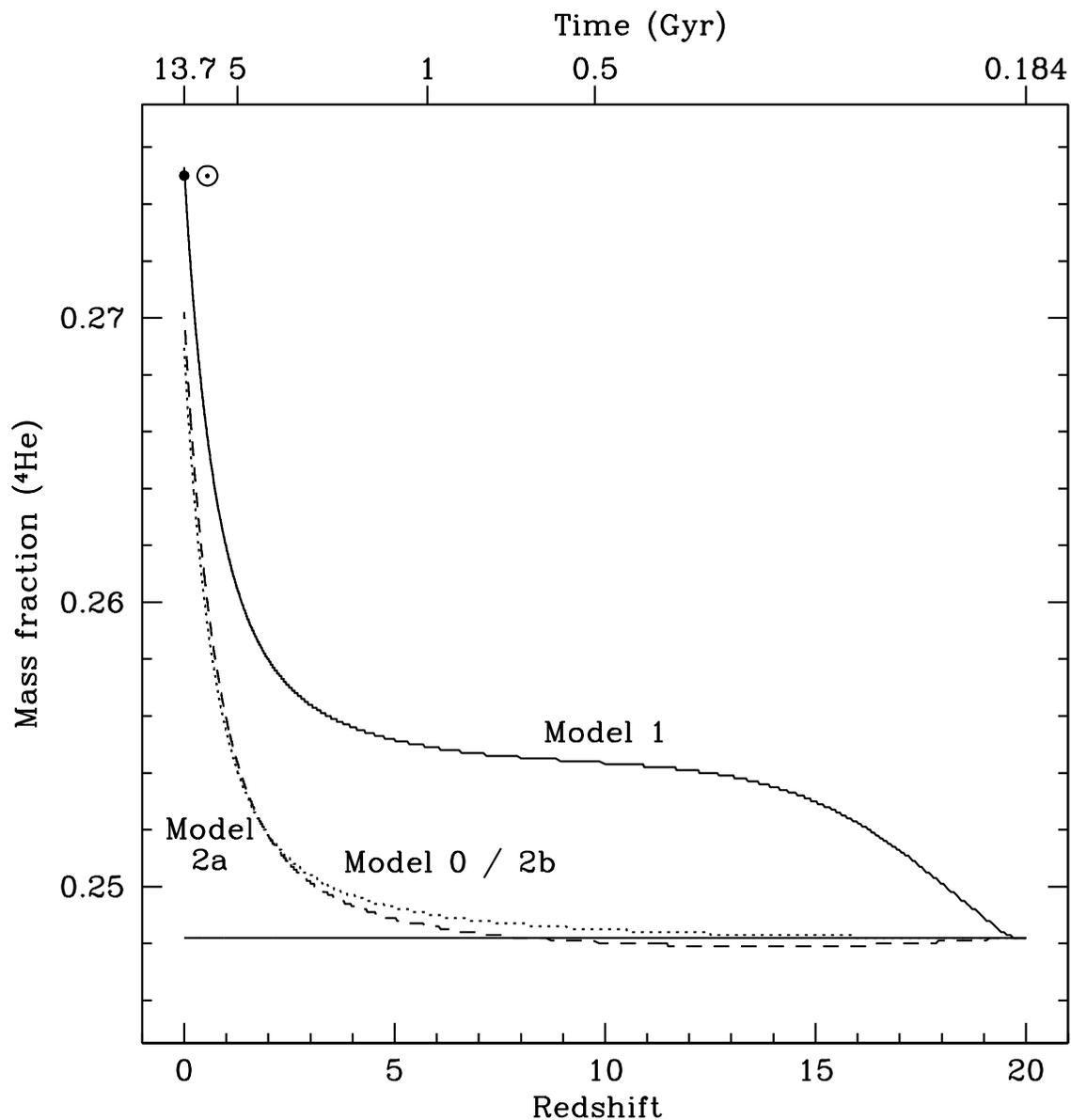}
\end{center}
\caption{\textbf{Evolution of the Helium abundance.} The mass fraction
of Helium in the cosmic structures is plotted as a function of
redshift for Model 0 and 2b (dotted line), Model 1 (solid line) and
Model 2a (dashed line). The BBN value is indicated as a horizontal
solid line, which in all models also corresponds to the value in the
IGM. The local abundance is indicated by the symbol $\odot$.}
\label{fig:He}
\end{figure}

\begin{figure}
\begin{center}
\plotone{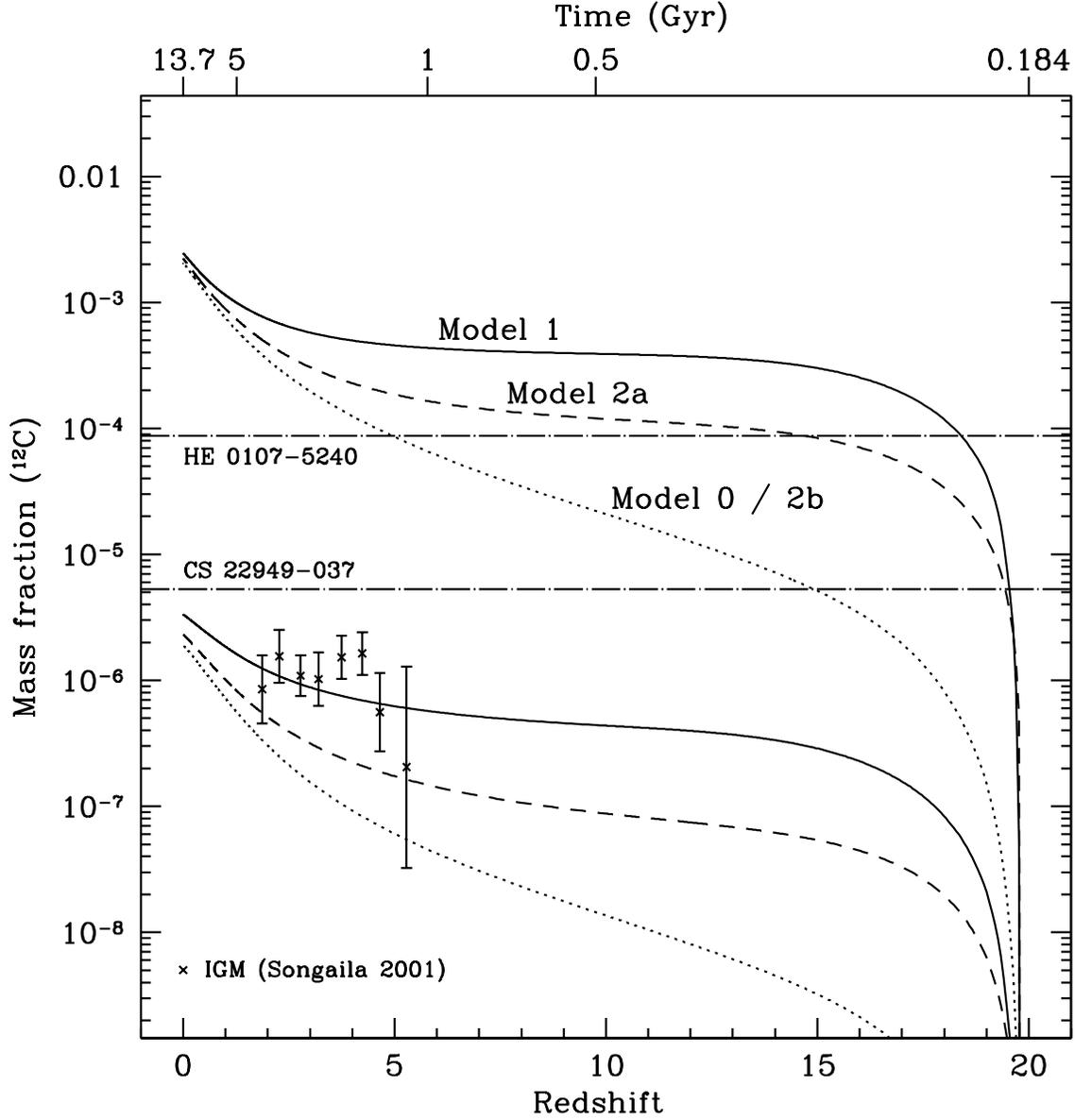}
\end{center}
\caption{\textbf{Evolution of the Carbon abundance.} The Carbon mass
fraction is plotted as a function of redshift for three models : Model
0 (dotted line), Model 1 (solid line) and Model 2a (dashed line) both
in the ISM of the cosmic structures (upper curves) and in the IGM
(lower curves). The predictions of Model 2b are exactly the same as in
Model 0. Note that data from \citet{songaila:01} in the IGM represent the abundance of CIV, which is a lower limit of the total Carbon abundance.} 
\label{fig:C}
\end{figure}

\begin{figure}
\begin{center}
\plotone{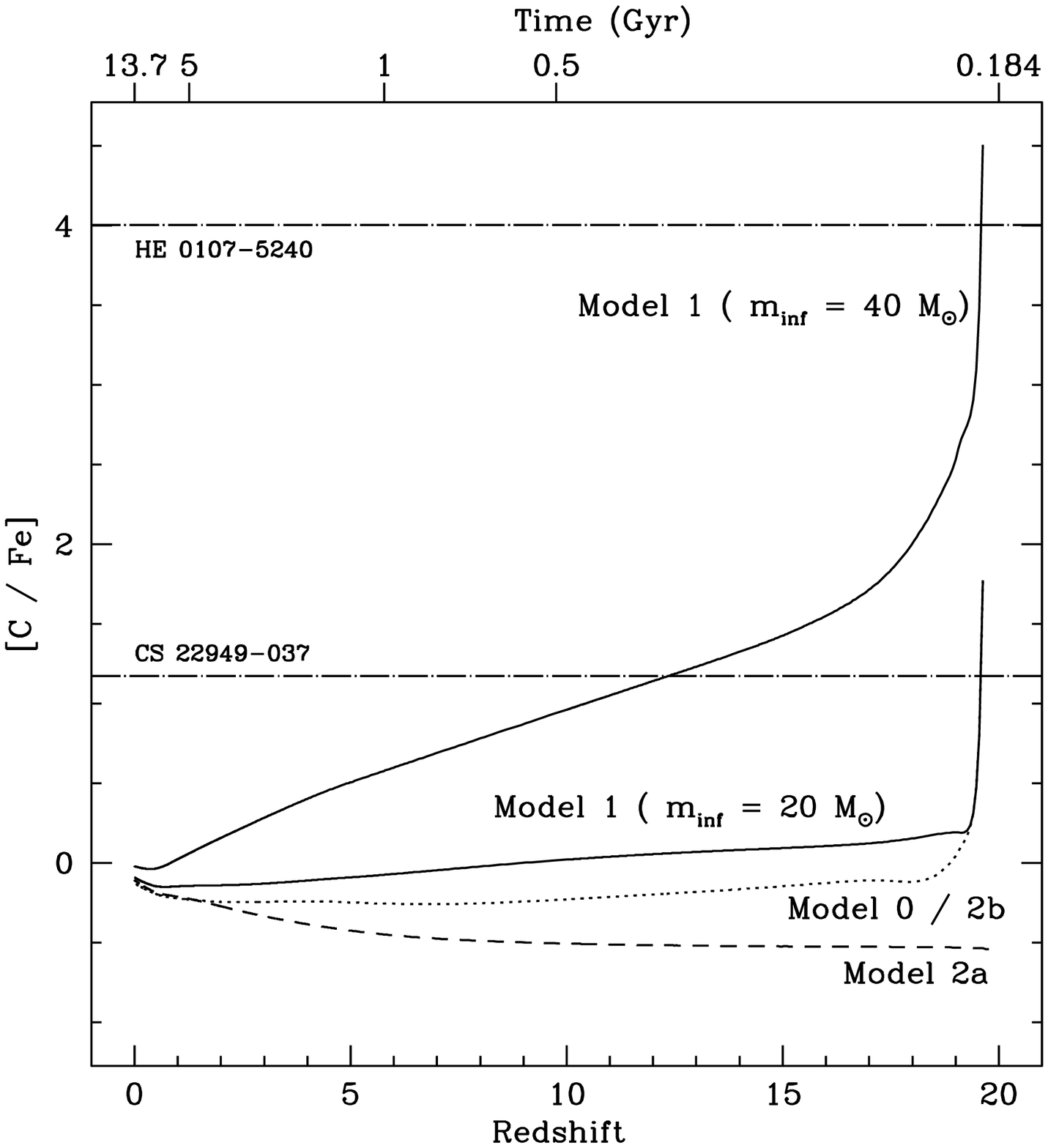}
\end{center}
\caption{\textbf{Evolution of the [C/Fe] ratio.} This ratio in the
cosmic structures is plotted as a function of redshift for Model 0 and
2b (dotted line), Model 1 (solid line) and Model 2a (dashed line). For
Model 1, two lower mass limits of the IMF are considered :
$m_\mathrm{inf}=20\ \mathrm{M}_{\odot}$ or $m_\mathrm{inf}=40\
\mathrm{M}_{\odot}$.}
\label{fig:CFe}
\end{figure}

\begin{figure}
\begin{center}
\plotone{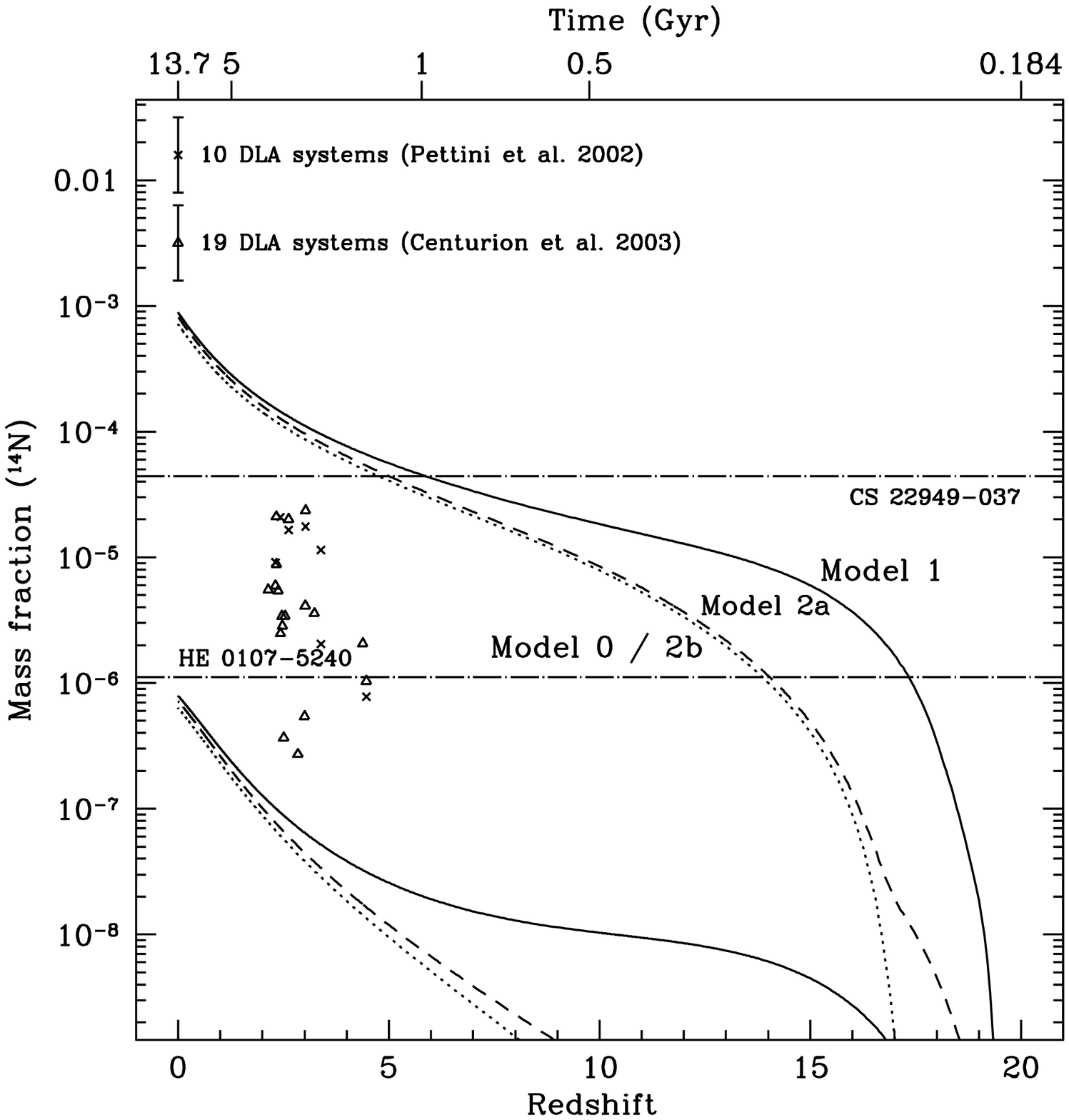}
\end{center}
\caption{\textbf{Evolution of the Nitrogen abundance.} The Nitrogen mass
fraction is plotted as a function of redshift as in
Figure~\ref{fig:C}.}
\label{fig:N}
\end{figure}

\begin{figure}
\begin{center}
\plotone{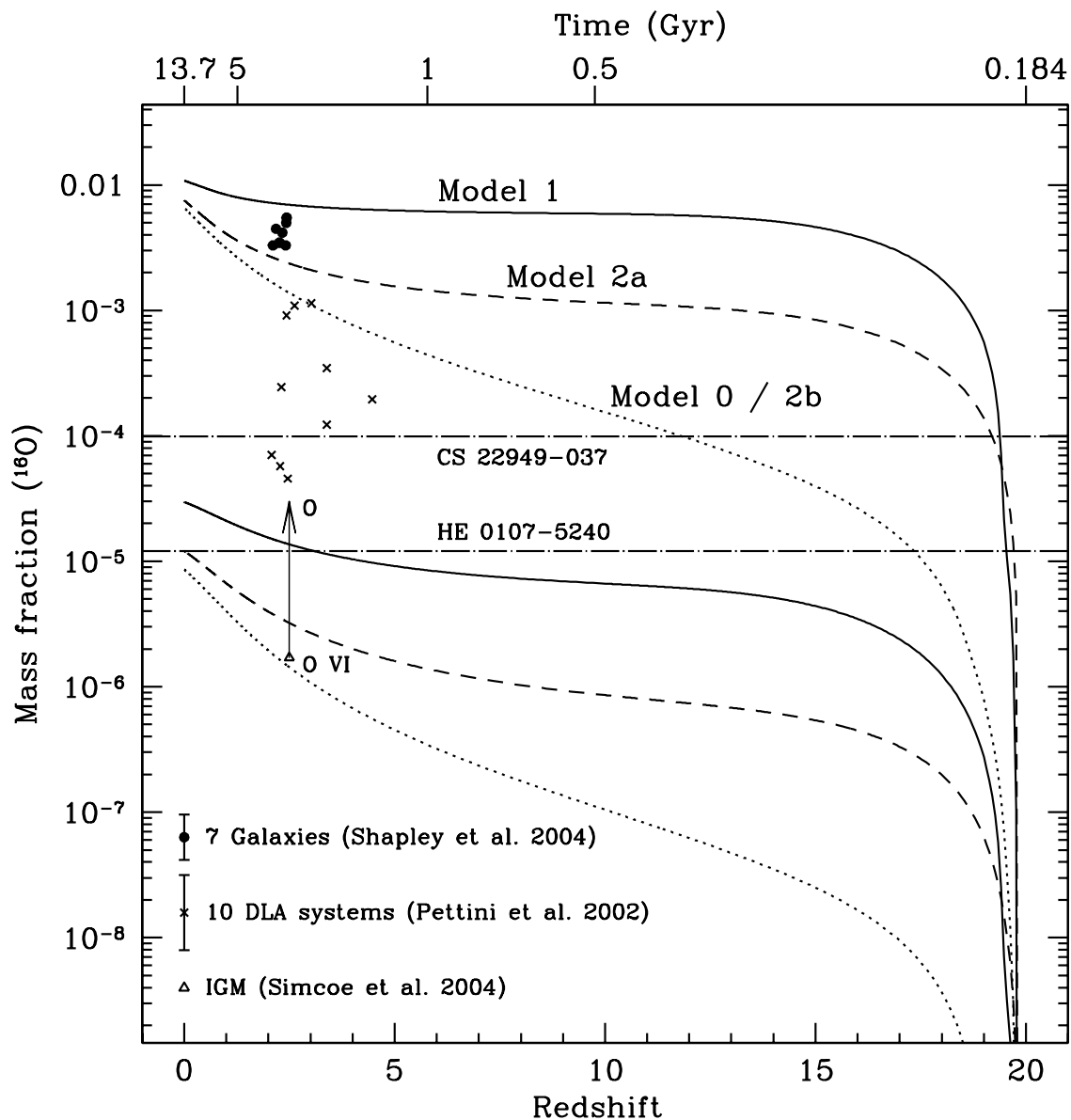}
\end{center}
\caption{\textbf{Evolution of the Oxygen abundance.} The Oxygen mass
fraction is plotted as a function of redshift as in
Figure~\ref{fig:C}. Note that \citet{simcoe:04} measure the abundance of O VI in the IGM. The arrow indicates the value of the Oxygen abundance they derive using ionization correction factors.}
\label{fig:O}
\end{figure}

\begin{figure}
\begin{center}
\plotone{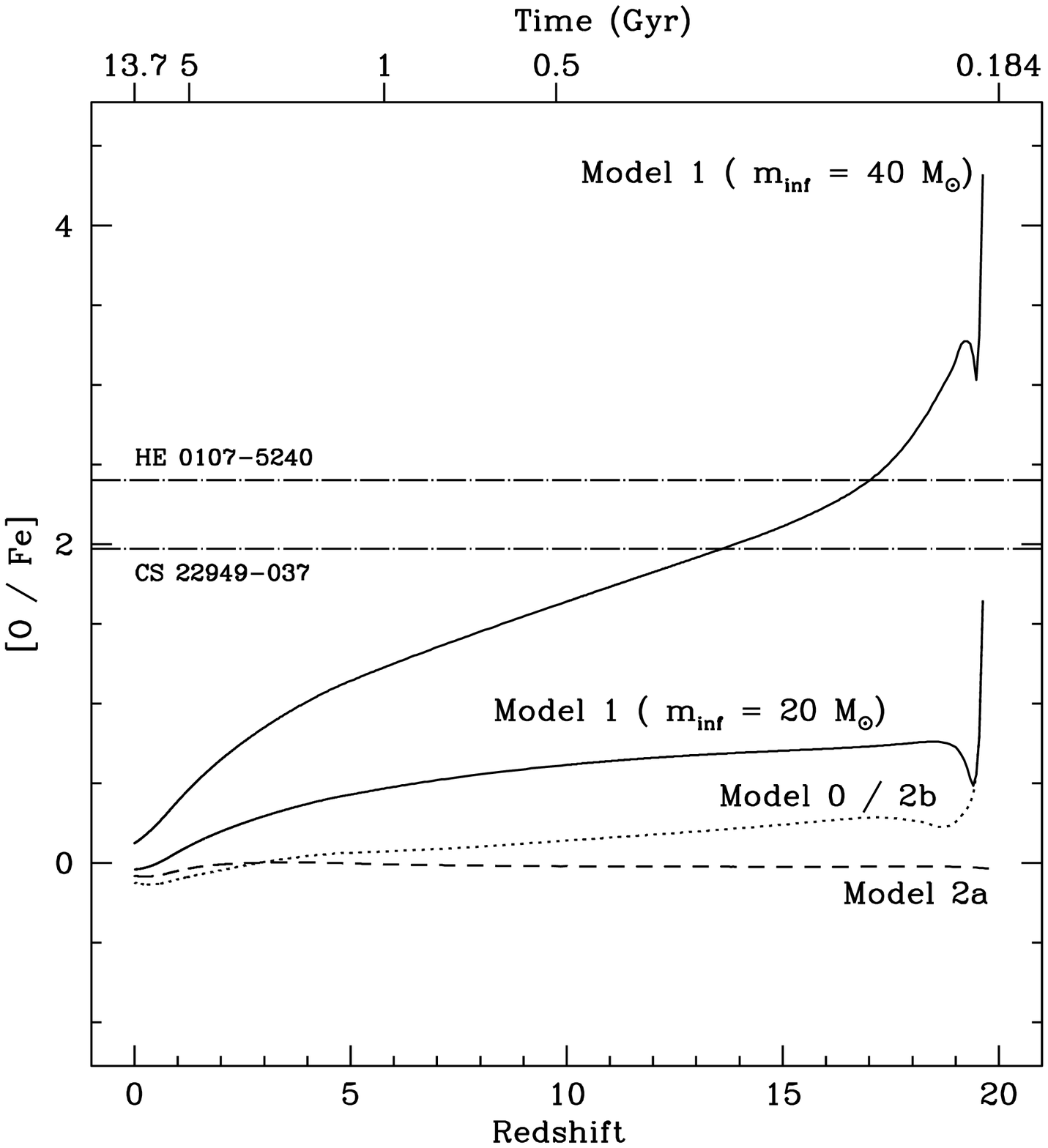}
\end{center}
\caption{\textbf{Evolution of the [O/Fe] ratio.} The Oxygen to Iron ratio in the
cosmic structures is plotted as a function of redshift for Model 0 and
2b (dotted line), Model 1 (solid line) and Model 2a (dashed line). For
Model 1, two lower mass limits of the IMF are considered :
$m_\mathrm{inf}=20\ \mathrm{M}_{\odot}$ or $m_\mathrm{inf}=40\
\mathrm{M}_{\odot}$.}
\label{fig:OFe}
\end{figure}

\begin{figure}
\begin{center}
\plotone{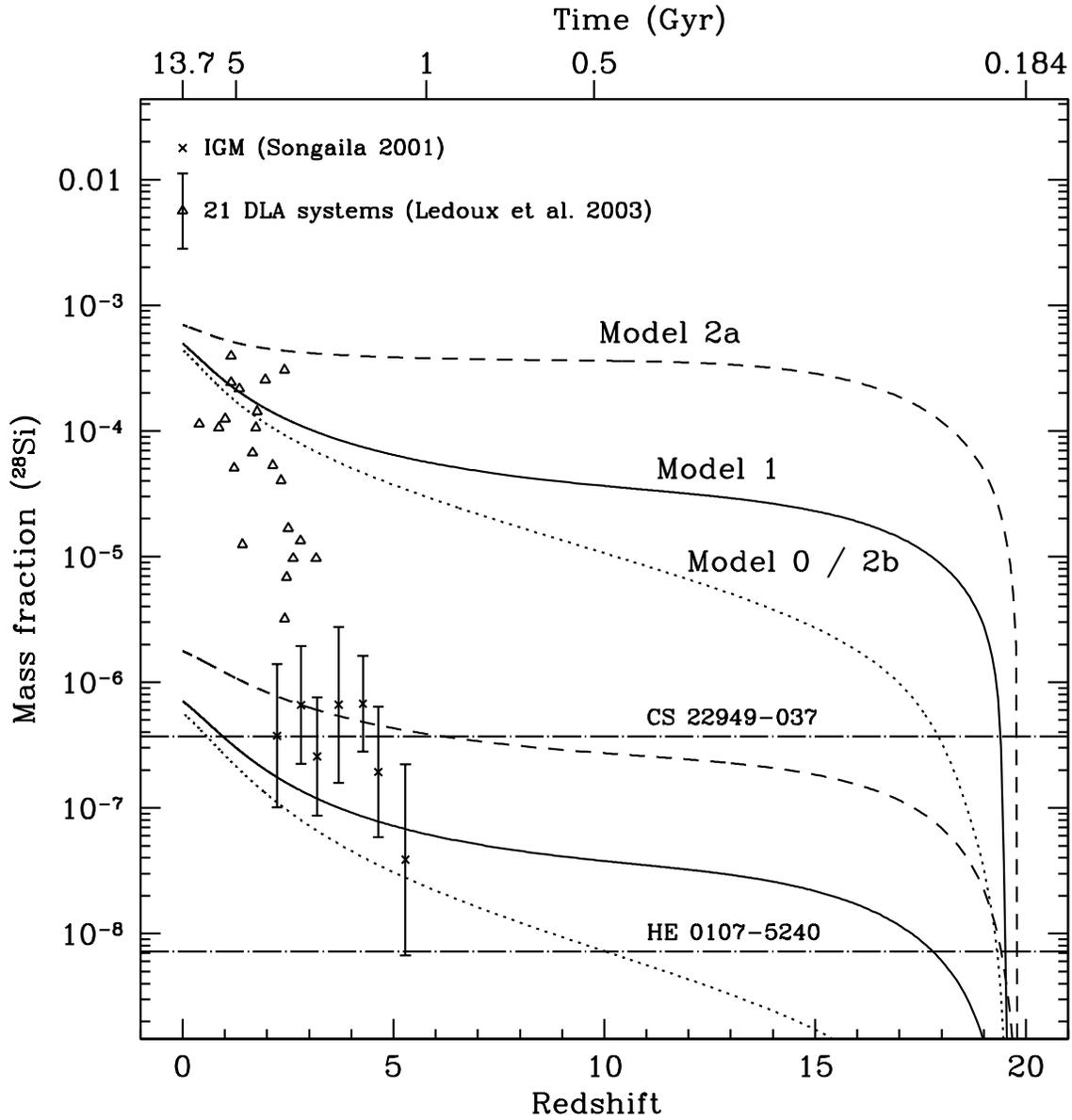}
\end{center}
\caption{\textbf{Evolution of the Silicon abundance.} The Silicon mass
fraction is plotted as a function of redshift as in
Figure~\ref{fig:C}.
Note that data from \citet{songaila:01} in the IGM represent the abundance of SiIV, which is a lower limit of the total Silicon abundance.}
\label{fig:Si}
\end{figure}

\begin{figure}
\begin{center}
\plotone{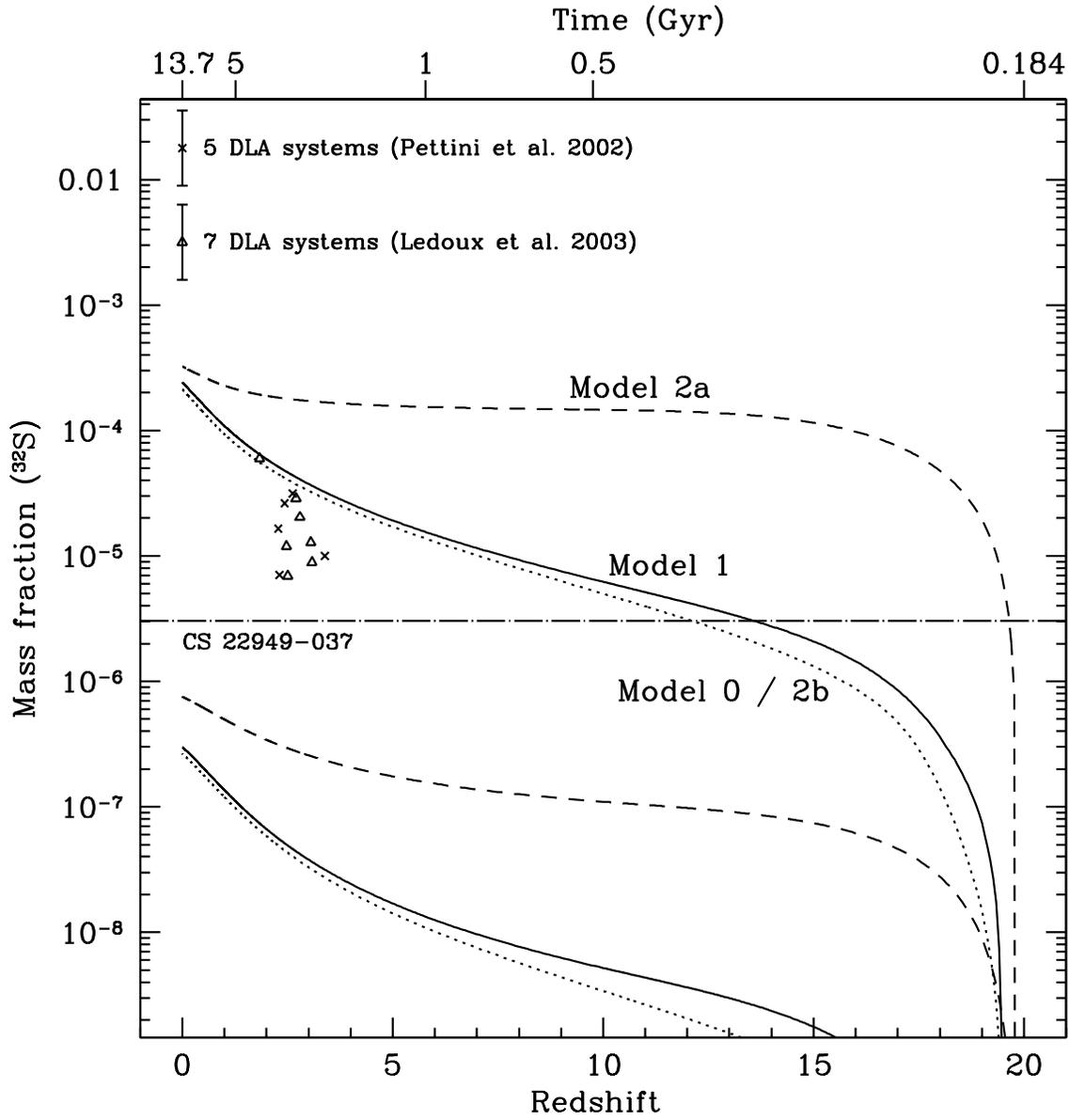}
\end{center}
\caption{\textbf{Evolution of the Sulfur abundance.} The Sulpher mass
fraction is plotted as a function of redshift as in
Figure~\ref{fig:C}.}
\label{fig:S}
\end{figure}

\begin{figure}
\begin{center}
\plotone{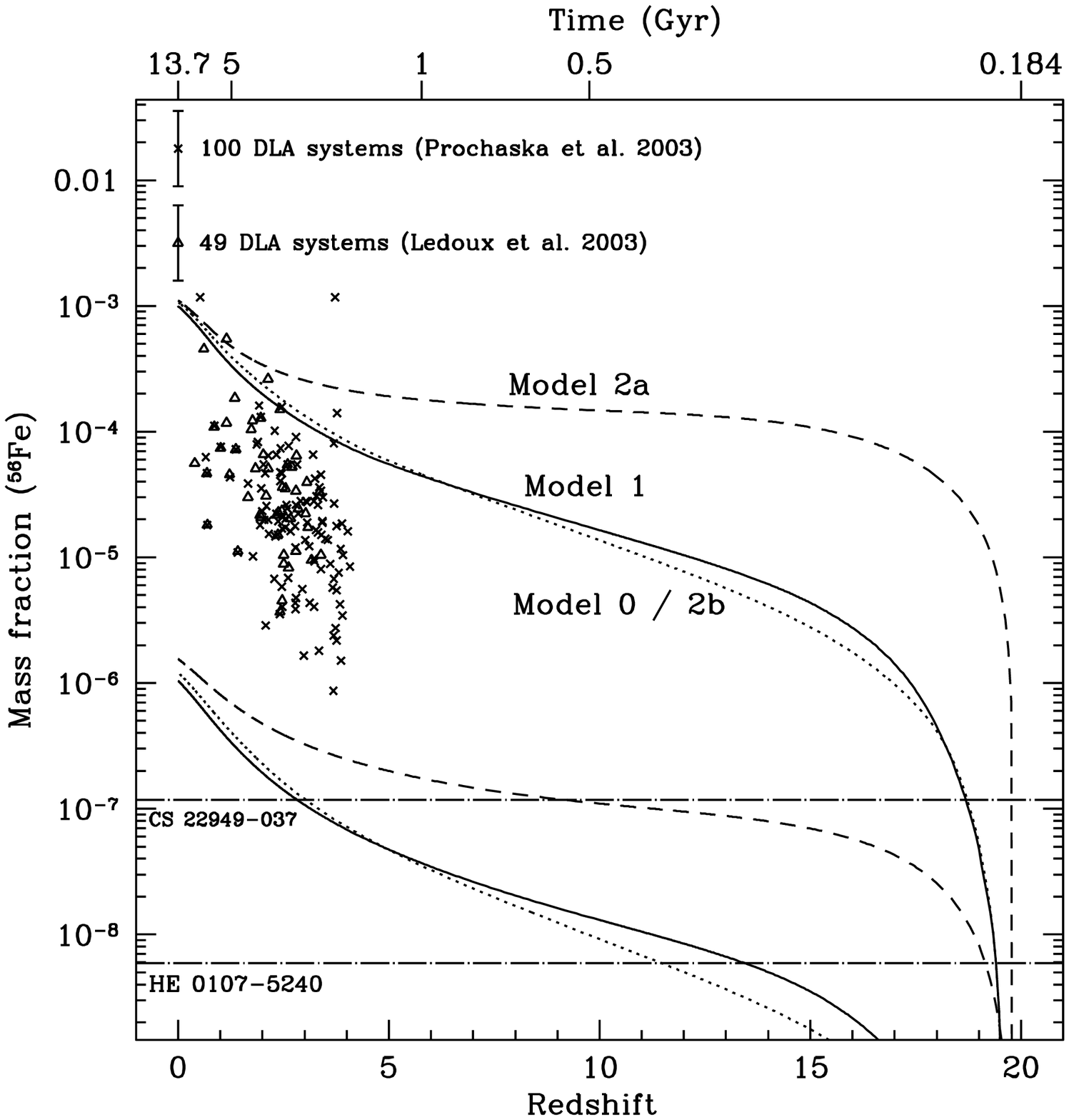}
\end{center}
\caption{\textbf{Evolution of the Iron abundance.} The Iron mass
fraction is plotted as a function of redshift as in
Figure~\ref{fig:C}.}
\label{fig:Fe}
\end{figure}

\begin{figure}
\begin{center}
\plotone{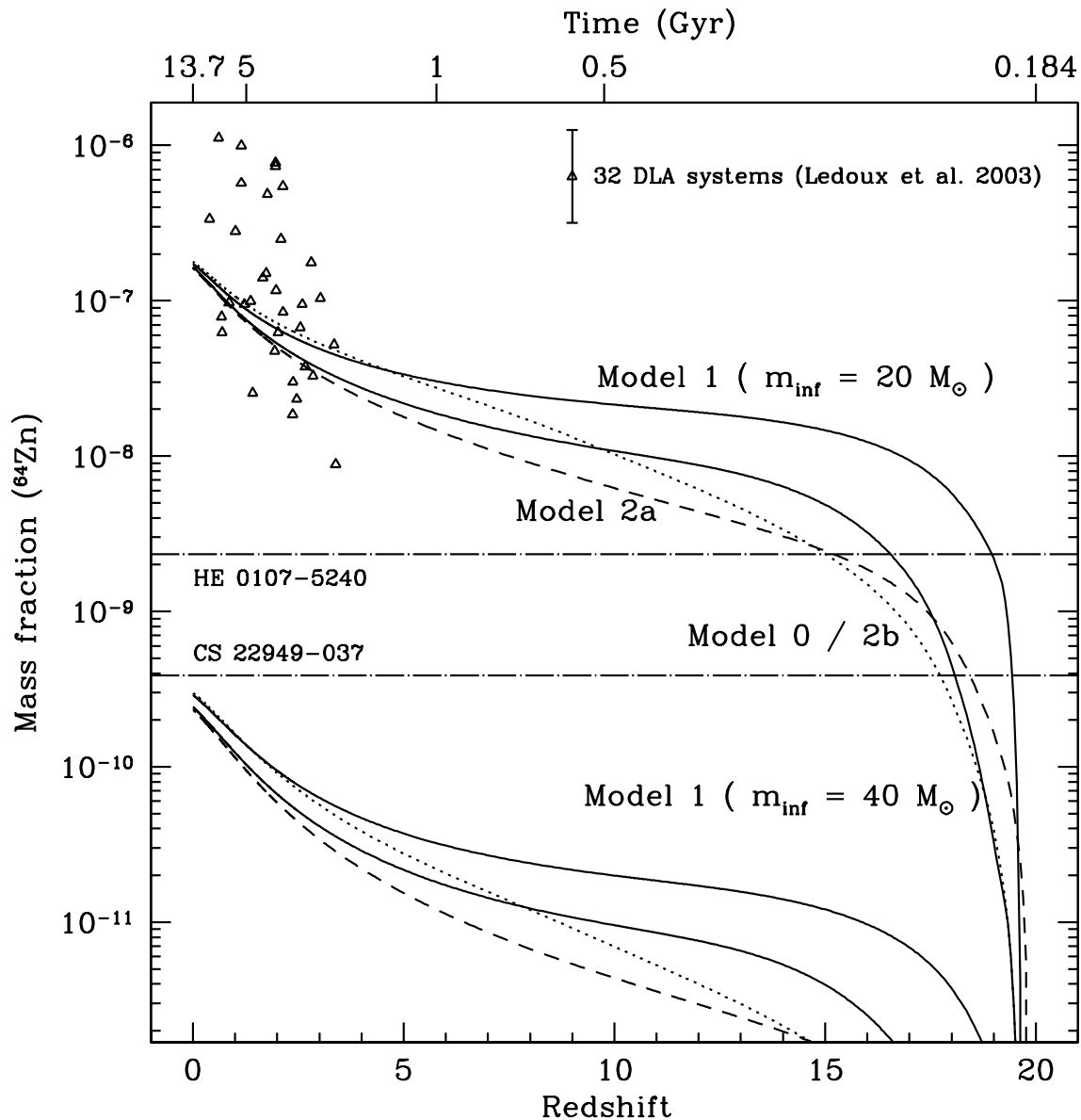}
\end{center}
\caption{\textbf{Evolution of the Zinc abundance.} The Zinc mass
fraction is plotted as a function of redshift as in
Figure~\ref{fig:C}. For Model 1, two lower mass limits of the
IMF are considered : $m_\mathrm{inf}=20\ \mathrm{M}_{\odot}$ or
$m_\mathrm{inf}=40\ \mathrm{M}_{\odot}$.}
\label{fig:Zn}
\end{figure}

\begin{figure}
\begin{center}
\plotone{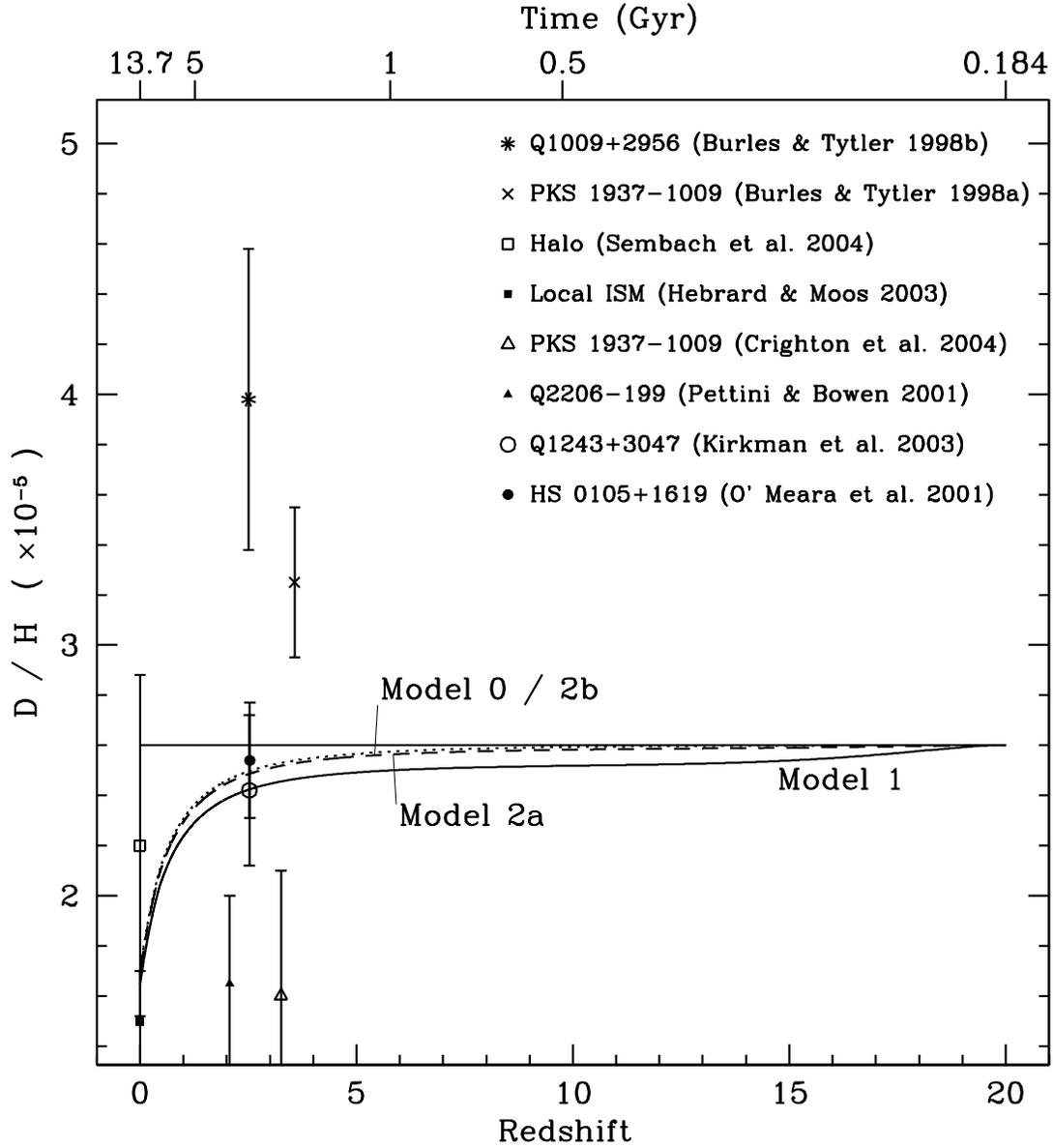}
\end{center}
\caption{\textbf{Evolution of the Deuterium abundance.} The $D/H$
ratio is plotted as a function of redshift in the ISM of the cosmic
structures for Model 0 and 2b (dotted line), Model 1 (solid line) and
Model 2a (dashed line). The BBN value is indicated as an horizontal
solid line, which in all models also corresponds to the value in the
IGM.}
\label{fig:D}
\end{figure}

In Figs.~\ref{fig:He} - \ref{fig:D}, we show the resulting evolution
of He, C, C/Fe, N, O, O/Fe, Si, S, Fe, Zn, and D. While there is no
high redshift data to compare the He abundance to, Figure~\ref{fig:He}
shows an interesting signature of Model 1 compared to the others\,: a
noticeable supplementary amount of Helium is produced at high redshift
by the massive stars.  The local value is however well fit. In Model
2a, only a small amount of Helium is produced because of the shift in
stellar nucleosynthesis towards the heavy elements.

Despite the fact that the late chemical evolution of Model 0 is quite
 acceptable, the observed abundances of the star HE 0107-5240, if
 taken as representative of the very early epoch under consideration,
 are not reproduced, especially the Carbon and Oxygen to Iron ratios,
 as seen in Figures~\ref{fig:CFe} and \ref{fig:OFe} where the observed
 level of these ratios is never attained. In addition the Carbon,
 Nitrogen, Oxygen and Sulfur abundances (Figures~\ref{fig:C},
 \ref{fig:N}, \ref{fig:O} and \ref{fig:S}) are poorly fit.  By
 that we mean that in this model, the abundance of Fe and Zn is
 reached at very high redshift ($z\sim 19$), as expected, whereas the
 abundance level of the other elements (C,N,O and S) is attained at a
 later epoch ($z\sim 12-16$).
A good fit to the abundance pattern of this star would require the
model to reach the observed abundance for each element at the same
redshift. Therefore it seems that Model 0 can also be rejected on
the basis of chemical criteria at high redshift.

Overall, the abundances predicted by Model 1 in the local structures
are consistent with observations. The late evolution of the
Silicon, Sulfur and Iron abundances in the cosmic structures
(Figures~\ref{fig:Si}, \ref{fig:S}, and \ref{fig:Fe}) is
also in good agreement with the observations in DLAs 
\citep{pettini:02,pettini:03,ledoux:03,prochaska:03} and the
late evolution of the Carbon and Silicon abundances in the IGM
(Figures~\ref{fig:C} and \ref{fig:Si}) is consistent with the
observations in the Lyman-$\alpha$ forest \citep{songaila:01}. 
Notice however that \citet{songaila:01} measures the abundance of \ion{C}{4} and \ion{Si}{4} and does not apply
any ionization correction. The work of \citet{schaye:03} and \citet{aguirre:04} indicates that the total
abundance of C and Si could be a factor of $\sim$ 5-7 larger. 
The case of
Oxygen is more problematic. 
The Oxygen abundance evolution predicted in  Model 1 is 
in good agreement with the recent observations in massive star forming galaxies 
at $z \ge 2$ \citep{shapley:04}. 
These observations clearly indicate little evolution of the
Oxygen abundance from $z=2$ to $z=0$, and favor an early enrichment at $z > 2$.
However, Figure~\ref{fig:O} shows that Oxygen is overabundant 
when compared to the observations in DLAs \citep{pettini:02}.
 One possible observational bias is related to the fact that the
Oxygen abundance can be measured only if the observed lines are not
saturated, which favors the weakest systems. On the other hand, a detailed
study of the impact of the uncertainty in stellar yields should also
be considered. Finally, Model 1 predictions in the IGM agree with the estimate
of the Oxygen abundance in the Lyman-$\alpha$ forest \citep{simcoe:04}. Note
that only O VI is directly measured by these authors. Therefore their estimate of the total
Oxygen abundance is limited by the uncertainties of the ionization state of Oxygen in these regions.

Moreover, if we adopt the extreme metal-poor halo stars HE 0107-5240
([Fe/H]=-5.3) and CS 22949-037 ([Fe/H]=-4) as tracers of the very
early Universe, it is remarkable that Model 1 is still globally
consistent with the absolute abundances of Carbon, Oxygen, Silicon,
Iron and Zinc (Figures~\ref{fig:C}, \ref{fig:O}, \ref{fig:Si},
\ref{fig:Fe} and \ref{fig:Zn}). The upper limit for the abundance of
Sulfur which is available for HE 0107-5240 does not constrain the
model. However the prediction of Model 1 is not in very good
agreement with the abundance of Sulfur in CS 22949-037
(Figure~\ref{fig:S}). That is, the build-up of S occurs late in the
model with respect to what is observed in this star. Similarly, the
abundance of Nitrogen is reproduced in a satisfactory manner in HE
0107-5240 but not in CS 22949-037 (Figure~\ref{fig:N}) due to the
extremely high abundance of N observed there.

We have also considered the predictions of Model 1 for the Carbon and
Oxygen to Iron ratios, as these ratios are very specific for these
extremely old halo-stars. The main quantity governing these ratios is
the lower mass of the stars formed in the initial massive
starburst. Figures~\ref{fig:CFe} and \ref{fig:OFe} compare the results
obtained for $m_\mathrm{inf}=20\ \mathrm{M}_{\odot}$ in comparison
with $m_\mathrm{inf}=40\ \mathrm{M}_{\odot}$ which has been used
elsewhere for Model 1.  As our calculations use the stellar yields of
\citet*{woosley:95} in this mass range, our results are directly
related to the evolution of the amount of Carbon, Oxygen and Iron
produced by stars from 20 M$_{\odot}$ to 40 M$_{\odot}$. In this
context, Figures~\ref{fig:CFe} and \ref{fig:OFe} show that
$m_\mathrm{inf}=40\ \mathrm{M}_{\odot}$ clearly favors high Carbon and
Oxygen to Iron ratios, as observed in both HE 0107-5240 and CS
22949-037. Conversely, $m_\mathrm{inf}=20\ \mathrm{M}_{\odot}$ favors
the early production of Zinc as this element is not synthesized for
$M>30\ \mathrm{M}_{\odot}$ at low metallicity according to \citet*{woosley:95}.
In this case, Figure~\ref{fig:Zn} shows a steeper
increase of the Zinc abundance between $z=20$ and $z=17$. 
 Finally, note that the local Zn abundance is not reproduced in any model,
 indicating the lack of Zn production in stellar models \citep{bihain:04}. 
 One can conclude that the observations of such stars provide strong
constraints for both the IMF of Pop III stars and stellar models,
 specifically  at
zero metallicity.

We conclude that Model 1 satisfies quite well all of the available
 constraints : when we superimpose an initial massive starburst to the
 normal mode, we predict a UV photon flux which is clearly sufficient
 to reionize the early Universe together with cosmic chemical
 evolution in the structures and in the IGM which is globally 
consistent with
 the local abundances and the abundances measured in DLAs and in the
 Lyman-$\alpha$ forest, as well as in the extreme metal-poor halo
 stars. In addition, we predict that stars like HE 0107-5240 probably trace
a generation of very old stars with a very specific chemical
 signature, and which were formed in the very first mini-halos and
 are thus well localized in time.

Next, we turn our attention back to Model 2a.  The local abundances
 (C, O, N, Si, S and Fe) are still well reproduced. Compared to
 Model 1, 
 the overabundance of Oxygen in the structures and in
 the IGM is reduced (see Figure~\ref{fig:O}) as was the case for
 He. However, Silicon, Sulfur and Iron are now a little over-abundant
 at late times (Figures~\ref{fig:Si}, \ref{fig:S} and
 \ref{fig:Fe}). The Zinc abundance is the same as in Model 1 at late
 times, showing a local underabundance. In the IGM, the Silicon
 abundance is better reproduced than in Model 1 but Carbon is
 underabundant (Figures~\ref{fig:C} and \ref{fig:Si}). In conclusion,
 the late chemical evolution is globally less satisfactory than in
 Model 1. For the peculiar case of the extreme metal-poor halo stars,
 Figures~\ref{fig:CFe} and \ref{fig:OFe} clearly prove that the
 predicted nucleosynthesis of PISN stars is unable to produce high
 Carbon and Oxygen to Iron ratios as observed in HE 0107-5240 and CS
 22949-037.

Finally, Model 2b is a very peculiar case as stars above 270
M$_{\odot}$ emit a large amount of ionizing photons but produce no
metals: the reionization process and the cosmic chemical evolution are
now de-correlated. The only constraint on the intensity of the initial
massive starburst comes from the observed cosmic SFR.  Thus the
chemical evolution is identical to that in Model 0 and therefore
reproduces well the observed late evolution but cannot satisfy the
constraints at high redshift coming from HE 0107-5240 and CS
22949-037.  In light of this result, one can envision the possibility
of a three-component model, based on Model 1 + a third massive
component in the 270 -- 500 M$_{\odot}$ range. The addition of this
component, the intensity of which is limited by the cosmic SFR
constraint, will only increase the early ionizing flux.

These results show clearly that observations do not support the hypothesis
of the existence of an early very massive stellar population. All of the observational 
constraints are better reproduced by Model 1 with a massive mode in the range 40--100 $\mathrm{M}_\odot$. Such a conclusion was recently obtained by \citet{venkatesan:04} 
in a completely different astrophysical context\,: high redshift quasars with solar of higher metal abundances.

Before concluding this section, we would like to address the question
of the evolution of D/H in the structures.  Unlike the evolution of
most other elements, the abundance of D/H decreases monotonically in
time.  The evolution of D/H in the models considered here is shown in
Fig.~\ref{fig:D}.  As one can see, in all models including Model 1,
there is very little change in D/H at early times, At lower redshift,
as D-free material is returned to the ISM, the D/H abundance begins to
drop.  In the figure, we compare the evolution of D/H to the deuterium
abundance observed in quasar absorption systems.  Previously, it was
noted \citep{casse:98} that there is a relation between the
amount of deuterium destruction and the rise in the observed cosmic
star formation rate.  In the models considered there, the timescale
over which the massive mode was operative was significantly longer
than that considered here. Because of the necessity to reionize the
Universe at very high redshift, the timescale associated with the
massive mode in Models 1 and 2 is considerably shorter, thereby
reducing the effect of the massive stars on the overall destruction of
D/H.

The average observed D/H abundance is in very good agreement with the BBN
predicted value of of D/H \citep{cyburt:03,coc:04} based on
the WMAP-inferred baryon density \citep{spergel:03}, however, there
appears to be considerable scatter in the data.  While the scatter may
be due to unresolved systematic uncertainties in the data, it may also
be a signature of specific chemical evolution.  In 
\citet{fields:01}, it was argued that significant changes in D/H could be
achieved if there is an enhanced population of intermediate mass
stars. The same model has also been used recently to explain
\citep*{ashenfelter:04a,ashenfelter:04b} the apparent observations of
variations in the fine-structure constant in the similar quasar
absorption systems \citep{murphy:03,chand:04}.  Although we
have not presented the results here, we have tested the idea that
intermediate mass stars ($\sim 3$--$8\ \mathrm{M}_{\odot}$) dominated
the IMF at $z \approx 20$ and have found that an enhanced population
of intermediate mass stars could not provide enough ionizing photons
to reionize the IGM.  Thus, such a population, if it existed, would
operate at later times after the prompt enrichment from massive stars
as in Models 1 and 2.

\section{Conclusions}

It is likely that ionizing photons from stars, as opposed to say
accreting black holes (in effect quasars or miniquasars)
are responsible for  the  early epoch of
reionization  at $z\sim 15-20$ inferred from the 
 WMAP results \citep[cf.][]{dijkstra:04a}.
However there are severe limitations on  massive stars if these formed
with
a ``normal'' IMF. The hypothesis of  a generation of very massive metal-free
stars helps supply the ionization requirements \citep*{cen:03a,wyithe:03}.
 However, we show here
 that chemical evolution constraints essentially rule out such an
 interpretation.
Rather, we describe how a standard massive IMF ($40$-$100\ \mathrm{M}_{\odot}$), 
for example as discussed by \citet{ciardi:03},
can  reionize the early intergalactic medium (see also a discussion of an even less massive IMF by 
\citealt*{venkatesan:03b}).

Our approach is to use the cosmic star formation rate history as our
primary guide to early star formation rates, and work with integrated
quantities in the context of hierarchical structure formation
to yield the ionizing photon flux history for 
alternative cosmic star formation models
which are capable of reionizing the early intergalactic medium.  We
have developed  models which include an early burst of massive stars combined
with standard star formation.  We computed the stellar ionizing flux of
photons and we tracked the nucleosynthetic yields for several
elements\,: D, $^{4}$He, C, N, O, Si, S, Fe, Zn as a function of redshift, both in the
IGM and in the ISM of the growing structures.  We compared the results of these models
with the observed abundances in the Lyman $\alpha$ forest and in 
DLAs.  We also considered
possible constraints associated with the observations of the two
extremely metal-poor stars HE 0107-5240 and CS22949-037.  We have shown 
that a bimodal
(or top-heavy) IMF (40 - 100 M$_\odot$) best satifies all constraints applied. 
In contrast, models with an extreme mass range ($\ga$ 100 M$_\odot$) often assumed
to be responsible for the early stages of reionization do less well.

As motivated by the numerical simulations of first star formation
\citep*{abel:02,bromm:02},
a  mode of even more extreme stellar masses in the
range ($\ge$ 270 M$_{\odot}$) has also been considered. As all massive
stars collapse entirely into black holes in this mode, the chemical
evolution and the reionization are de-correlated, as already mentioned by 
\citet{tumlinson:04}. The ionizing flux
from these very massive stars can easily reionize the Universe at
$z\sim 17$. However the chemical evolution in this case is exactly the
same as in the standard star formation model, and the high redshift
abundances are not reproduced. Consequently, the suggestion 
\citep*{bromm:03} that such Population III stars were the
precursors of the extreme metal-poor halo stars is untenable.
There is no evidence, nor any need, for a hypothesised 
primordial population of  very 
massive 
stars in order to account  for  the 
chemical abundances of extremely metal-poor halo stars
or of the intergalactic medium.
The combined population of early-forming, 
normal (0.1 - 100 M$_\odot$) and massive (40 - 100 M$_\odot$) stars
can simultaneously explain 
the cosmic chemical evolution and the observations of
extremely metal-poor halo stars
and also account for early 
cosmological reionization.

%

We have shown that  the initial massive starburst, which
originally was introduced to reionize the early Universe, produces
rapid initial metal pollution. The existence of old, C-rich halo stars
with high [O/Fe] and [C/Fe] ratios is predicted as a consequence of
these massive stars. The recently observed abundances in the oldest
halo stars could trace this very specific stellar population.

We have also found that the D
abundance is strongly coupled to the gas fraction in the structures,
with the implication that local D measurements are a non-robust
cosmological probe. In addition,  there is some non-primordial contribution to
the He abundance even in metal-poor galaxies, that is however within
the observed range.

Our suggestion is far from the whole story. There must be late
ionization input by harder photons than produced by OB stars to
account for the \ion{He}{2} abundance at $z\sim 2-4,$
for example associated with the quasar population 
\citep*{hui:03} or  Pop III stars \citep*{bromm:01a,venkatesan:03a}.
As structure builds up,
the ionizing photon escape fraction will surely decrease,
and the  intergalactic medium is likely to recombine before 
$z\sim 6,$  at which time the neutral fraction is constrained from the onset
of the Gunn-Peterson effect to be about 0.001\%. 
\citet{cen:03b} has suggested that the Universe becomes neutral again at $z \sim 13$ and is reionized for the second time at $z\sim 6$, not necessary by the same stellar population. However our calculations suggests that
normal star formation
is likely to suffice at this epoch \citep{gnedin:04} in order to account for the observed
ionization level.

\acknowledgments The authors warmly thank  Roger Cayrel and Patrick
Petitjean for frequent fruitful discussions.  This work was supported
in part by DOE grants DE-FG02-94ER-40823 at the University of
Minnesota, and by PICS 1076 CNRS France/USA.

\citetalias{burles:98a}
\citetalias{burles:98b}
\citetalias{crighton:04}
\citetalias{hebrard:03}
\citetalias{kirkman:03}
\citetalias{omeara:01}
\citetalias{pettini:01}
\citetalias{sembach:04}
\citetalias{centurion:03}

\bibliographystyle{apj}
\bibliography{ms}

\begin{thebibliography}{90}
\expandafter\ifx\csname natexlab\endcsname\relax\def\natexlab#1{#1}\fi

\bibitem[{{Abel} {et~al.}(2002){Abel}, {Bryan}, \& {Norman}}]{abel:02}
{Abel}, T., {Bryan}, G.~L., \& {Norman}, M.~L. 2002, Science, 295, 93

\bibitem[{{Aguirre} {et~al.}(2004){Aguirre}, {Schaye}, {Kim}, {Theuns},
  {Rauch}, \& {Sargent}}]{aguirre:04}
{Aguirre}, A., {Schaye}, J., {Kim}, T., {Theuns}, T., {Rauch}, M., \&
  {Sargent}, W.~L.~W. 2004, \apj, 602, 38

\bibitem[{{Ashenfelter} {et~al.}(2004{\natexlab{a}}){Ashenfelter}, {Mathews},
  \& {Olive}}]{ashenfelter:04a}
{Ashenfelter}, T., {Mathews}, G.~J., \& {Olive}, K.~A. 2004{\natexlab{a}},
  \prl, 92, 041102

\bibitem[{{Ashenfelter} {et~al.}(2004{\natexlab{b}}){Ashenfelter}, {Mathews},
  \& {Olive}}]{ashenfelter:04b}
{Ashenfelter}, T.~P., {Mathews}, G.~J., \& {Olive}, K.~A. 2004{\natexlab{b}},
  astro-ph/0404257

\bibitem[{{Becker} {et~al.}(2001){Becker}, {Fan}, {White}, {Strauss},
  {Narayanan}, {Lupton}, {Gunn}, {Annis}, {Bahcall}, {Brinkmann}, {Connolly},
  {Csabai}, {Czarapata}, {Doi}, {Heckman}, {Hennessy}, {Ivezi{\' c}}, {Knapp},
  {Lamb}, {McKay}, {Munn}, {Nash}, {Nichol}, {Pier}, {Richards}, {Schneider},
  {Stoughton}, {Szalay}, {Thakar}, \& {York}}]{becker:01}
{Becker}, R.~H., {Fan}, X., {White}, R.~L., {Strauss}, M.~A., {Narayanan},
  V.~K., {Lupton}, R.~H., {Gunn}, J.~E., {Annis}, J., {Bahcall}, N.~A.,
  {Brinkmann}, J., {Connolly}, A.~J., {Csabai}, I., {Czarapata}, P.~C., {Doi},
  M., {Heckman}, T.~M., {Hennessy}, G.~S., {Ivezi{\' c}}, {\v Z}., {Knapp},
  G.~R., {Lamb}, D.~Q., {McKay}, T.~A., {Munn}, J.~A., {Nash}, T., {Nichol},
  R., {Pier}, J.~R., {Richards}, G.~T., {Schneider}, D.~P., {Stoughton}, C.,
  {Szalay}, A.~S., {Thakar}, A.~R., \& {York}, D.~G. 2001, \aj, 122, 2850

\bibitem[{{Bessell} {et~al.}(2004){Bessell}, {Christlieb}, \&
  {Gustafsson}}]{bessell:04}
{Bessell}, M.~S., {Christlieb}, N., \& {Gustafsson}, B. 2004, astro-ph/0401450

\bibitem[{{Bihain} {et~al.}(2004){Bihain}, {Israelian}, {Rebolo}, {Bonifacio},
  \& {Molaro}}]{bihain:04}
{Bihain}, G., {Israelian}, G., {Rebolo}, R., {Bonifacio}, P., \& {Molaro}, P.
  2004, astro-ph/0405050

\bibitem[{{Bromm}(2004)}]{bromm:04}
{Bromm}, V. 2004, \pasp, 116, 103

\bibitem[{{Bromm} {et~al.}(2002){Bromm}, {Coppi}, \& {Larson}}]{bromm:02}
{Bromm}, V., {Coppi}, P.~S., \& {Larson}, R.~B. 2002, \apj, 564, 23

\bibitem[{{Bromm} {et~al.}(2001{\natexlab{a}}){Bromm}, {Ferrara}, {Coppi}, \&
  {Larson}}]{bromm:01b}
{Bromm}, V., {Ferrara}, A., {Coppi}, P.~S., \& {Larson}, R.~B.
  2001{\natexlab{a}}, \mnras, 328, 969

\bibitem[{{Bromm} {et~al.}(2001{\natexlab{b}}){Bromm}, {Kudritzki}, \&
  {Loeb}}]{bromm:01a}
{Bromm}, V., {Kudritzki}, R.~P., \& {Loeb}, A. 2001{\natexlab{b}}, \apj, 552,
  464

\bibitem[{{Bromm} \& {Loeb}(2003)}]{bromm:03}
{Bromm}, V. \& {Loeb}, A. 2003, \nat, 425, 812

\bibitem[{{Burles} \& {Tytler}(1998{\natexlab{a}})}]{burles:98a}
{Burles}, S. \& {Tytler}, D. 1998{\natexlab{a}}, \apj, 499, 699

\bibitem[{{Burles} \& {Tytler}(1998{\natexlab{b}})}]{burles:98b}
---. 1998{\natexlab{b}}, \apj, 507, 732

\bibitem[{{Cass\'{e}} {et~al.}(1998){Cass\'{e}}, {Olive}, {Vangioni-Flam}, \&
  {Audouze}}]{casse:98}
{Cass\'{e}}, M., {Olive}, K.~A., {Vangioni-Flam}, E., \& {Audouze}, J. 1998,
  \na, 3, 259

\bibitem[{{Cen}(2003{\natexlab{a}})}]{cen:03a}
{Cen}, R. 2003{\natexlab{a}}, \apjl, 591, L5

\bibitem[{{Cen}(2003{\natexlab{b}})}]{cen:03b}
---. 2003{\natexlab{b}}, \apj, 591, 12

\bibitem[{{Centuri{\' o}n} {et~al.}(2003){Centuri{\' o}n}, {Molaro}, {Vladilo},
  {P{\' e}roux}, {Levshakov}, \& {D'Odorico}}]{centurion:03}
{Centuri{\' o}n}, M., {Molaro}, P., {Vladilo}, G., {P{\' e}roux}, C.,
  {Levshakov}, S.~A., \& {D'Odorico}, V. 2003, \aap, 403, 55

\bibitem[{{Chand} {et~al.}(2004){Chand}, {Srianand}, {Petitjean}, \&
  {Aracil}}]{chand:04}
{Chand}, H., {Srianand}, R., {Petitjean}, P., \& {Aracil}, B. 2004,
  astro-ph/0401094

\bibitem[{{Christlieb} {et~al.}(2004){Christlieb}, {Gustafsson}, {Korn},
  {Barklem}, {Beers}, {Bessell}, {Karlsson}, \&
  {Mizuno-Wiedner}}]{christlieb:04}
{Christlieb}, N., {Gustafsson}, B., {Korn}, A.~J., {Barklem}, P.~S., {Beers},
  T.~C., {Bessell}, M.~S., {Karlsson}, T., \& {Mizuno-Wiedner}, M. 2004, \apj,
  603, 708

\bibitem[{{Ciardi} {et~al.}(2003){Ciardi}, {Ferrara}, \& {White}}]{ciardi:03}
{Ciardi}, B., {Ferrara}, A., \& {White}, S.~D.~M. 2003, \mnras, 344, L7

\bibitem[{{Coc} {et~al.}(2004){Coc}, {Vangioni-Flam}, {Descouvemont},
  {Adahchour}, \& {Angulo}}]{coc:04}
{Coc}, A., {Vangioni-Flam}, E., {Descouvemont}, P., {Adahchour}, A., \&
  {Angulo}, C. 2004, \apj, 600, 544

\bibitem[{{Crighton} {et~al.}(2004){Crighton}, {Webb}, {Ortiz-Gil}, \&
  {Fernandez-Soto}}]{crighton:04}
{Crighton}, N.~H.~M., {Webb}, J.~K., {Ortiz-Gil}, A., \& {Fernandez-Soto}, A.
  2004, astro-ph/0403512

\bibitem[{{Cyburt} {et~al.}(2003){Cyburt}, {Fields}, \& {Olive}}]{cyburt:03}
{Cyburt}, R.~H., {Fields}, B.~D., \& {Olive}, K.~A. 2003, Physics Letters B,
  567, 227

\bibitem[{{Depagne} {et~al.}(2002){Depagne}, {Hill}, {Spite}, {Spite}, {Plez},
  {Beers}, {Barbuy}, {Cayrel}, {Andersen}, {Bonifacio}, {Fran{\c c}ois},
  {Nordstr{\" o}m}, \& {Primas}}]{depagne:02}
{Depagne}, E., {Hill}, V., {Spite}, M., {Spite}, F., {Plez}, B., {Beers},
  T.~C., {Barbuy}, B., {Cayrel}, R., {Andersen}, J., {Bonifacio}, P., {Fran{\c
  c}ois}, P., {Nordstr{\" o}m}, B., \& {Primas}, F. 2002, \aap, 390, 187

\bibitem[{{Dickinson} {et~al.}(2003){Dickinson}, {Papovich}, {Ferguson}, \&
  {Budav{\' a}ri}}]{dickinson:03}
{Dickinson}, M., {Papovich}, C., {Ferguson}, H.~C., \& {Budav{\' a}ri}, T.
  2003, \apj, 587, 25

\bibitem[{{Dijkstra} {et~al.}(2004{\natexlab{a}}){Dijkstra}, {Haiman}, \&
  {Loeb}}]{dijkstra:04b}
{Dijkstra}, M., {Haiman}, Z., \& {Loeb}, A. 2004{\natexlab{a}},
  astro-ph/0403078

\bibitem[{{Dijkstra} {et~al.}(2004{\natexlab{b}}){Dijkstra}, {Haiman}, {Rees},
  \& {Weinberg}}]{dijkstra:04a}
{Dijkstra}, M., {Haiman}, Z., {Rees}, M.~J., \& {Weinberg}, D.~H.
  2004{\natexlab{b}}, \apj, 601, 666

\bibitem[{{Fields} {et~al.}(2001){Fields}, {Olive}, {Silk}, {Cass{\' e}}, \&
  {Vangioni-Flam}}]{fields:01}
{Fields}, B.~D., {Olive}, K.~A., {Silk}, J., {Cass{\' e}}, M., \&
  {Vangioni-Flam}, E. 2001, \apj, 563, 653

\bibitem[{{Fukugita} {et~al.}(1998){Fukugita}, {Hogan}, \&
  {Peebles}}]{fukugita:98}
{Fukugita}, M., {Hogan}, C.~J., \& {Peebles}, P.~J.~E. 1998, \apj, 503, 518

\bibitem[{{Giavalisco} {et~al.}(2004){Giavalisco}, {Dickinson}, {Ferguson},
  {Ravindranath}, {Kretchmer}, {Moustakas}, {Madau}, {Fall}, {Gardner},
  {Livio}, {Papovich}, {Renzini}, {Spinrad}, {Stern}, \&
  {Riess}}]{giavalisco:04}
{Giavalisco}, M., {Dickinson}, M., {Ferguson}, H.~C., {Ravindranath}, S.,
  {Kretchmer}, C., {Moustakas}, L.~A., {Madau}, P., {Fall}, S.~M., {Gardner},
  J.~P., {Livio}, M., {Papovich}, C., {Renzini}, A., {Spinrad}, H., {Stern},
  D., \& {Riess}, A. 2004, \apjl, 600, L103

\bibitem[{{Gnedin}(2004)}]{gnedin:04}
{Gnedin}, N.~Y. 2004, astro-ph/0403699

\bibitem[{{H{\' e}brard} \& {Moos}(2003)}]{hebrard:03}
{H{\' e}brard}, G. \& {Moos}, H.~W. 2003, \apj, 599, 297

\bibitem[{{Haiman} \& {Holder}(2003)}]{haiman:03}
{Haiman}, Z. \& {Holder}, G.~P. 2003, \apj, 595, 1

\bibitem[{{Heger} {et~al.}(2003){Heger}, {Fryer}, {Woosley}, {Langer}, \&
  {Hartmann}}]{heger:03}
{Heger}, A., {Fryer}, C.~L., {Woosley}, S.~E., {Langer}, N., \& {Hartmann},
  D.~H. 2003, \apj, 591, 288

\bibitem[{{Heger} \& {Woosley}(2002)}]{heger:02}
{Heger}, A. \& {Woosley}, S.~E. 2002, \apj, 567, 532

\bibitem[{{Hui} \& {Haiman}(2003)}]{hui:03}
{Hui}, L. \& {Haiman}, Z. 2003, \apj, 596, 9

\bibitem[{{Israelian} {et~al.}(2004){Israelian}, {Shchukina}, {Rebolo},
  {Basri}, {Gonzalez-Hernandez}, \& {Kajino}}]{israelian:04}
{Israelian}, G., {Shchukina}, N., {Rebolo}, R., {Basri}, G.,
  {Gonzalez-Hernandez}, J.~I., \& {Kajino}, T. 2004, astro-ph/0403033

\bibitem[{{Iwata} {et~al.}(2003){Iwata}, {Ohta}, {Tamura}, {Ando}, {Wada},
  {Watanabe}, {Akiyama}, \& {Aoki}}]{iwata:03}
{Iwata}, I., {Ohta}, K., {Tamura}, N., {Ando}, M., {Wada}, S., {Watanabe}, C.,
  {Akiyama}, M., \& {Aoki}, K. 2003, \pasj, 55, 415

\bibitem[{{Kirkman} {et~al.}(2003){Kirkman}, {Tytler}, {Suzuki}, {O'Meara}, \&
  {Lubin}}]{kirkman:03}
{Kirkman}, D., {Tytler}, D., {Suzuki}, N., {O'Meara}, J.~M., \& {Lubin}, D.
  2003, \apjs, 149, 1

\bibitem[{{Kogut} {et~al.}(2003){Kogut}, {Spergel}, {Barnes}, {Bennett},
  {Halpern}, {Hinshaw}, {Jarosik}, {Limon}, {Meyer}, {Page}, {Tucker},
  {Wollack}, \& {Wright}}]{kogut:03}
{Kogut}, A., {Spergel}, D.~N., {Barnes}, C., {Bennett}, C.~L., {Halpern}, M.,
  {Hinshaw}, G., {Jarosik}, N., {Limon}, M., {Meyer}, S.~S., {Page}, L.,
  {Tucker}, G.~S., {Wollack}, E., \& {Wright}, E.~L. 2003, \apjs, 148, 161

\bibitem[{{Lanzetta} {et~al.}(2002){Lanzetta}, {Yahata}, {Pascarelle}, {Chen},
  \& {Fern{\' a}ndez-Soto}}]{lanzetta:02}
{Lanzetta}, K.~M., {Yahata}, N., {Pascarelle}, S., {Chen}, H., \& {Fern{\'
  a}ndez-Soto}, A. 2002, \apj, 570, 492

\bibitem[{{Larson}(1986)}]{larson:86}
{Larson}, R.~B. 1986, \mnras, 218, 409

\bibitem[{{Ledoux} {et~al.}(2003){Ledoux}, {Petitjean}, \&
  {Srianand}}]{ledoux:03}
{Ledoux}, C., {Petitjean}, P., \& {Srianand}, R. 2003, \mnras, 346, 209

\bibitem[{{Lilly} {et~al.}(1996){Lilly}, {Le Fevre}, {Hammer}, \&
  {Crampton}}]{lilly:96}
{Lilly}, S.~J., {Le Fevre}, O., {Hammer}, F., \& {Crampton}, D. 1996, \apjl,
  460, L1

\bibitem[{{Maeder} \& {Meynet}(1989)}]{maeder:89}
{Maeder}, A. \& {Meynet}, G. 1989, \aap, 210, 155

\bibitem[{{Mo} \& {White}(2002)}]{mo:02}
{Mo}, H.~J. \& {White}, S.~D.~M. 2002, \mnras, 336, 112

\bibitem[{{Murphy} {et~al.}(2003){Murphy}, {Webb}, \& {Flambaum}}]{murphy:03}
{Murphy}, M.~T., {Webb}, J.~K., \& {Flambaum}, V.~V. 2003, \mnras, 345, 609

\bibitem[{{Nagamine} {et~al.}(2003){Nagamine}, {Cen}, {Hernquist}, {Ostriker},
  \& {Springel}}]{nagamine:03}
{Nagamine}, K., {Cen}, R., {Hernquist}, L., {Ostriker}, J.~P., \& {Springel},
  V. 2003, astro-ph/0311294

\bibitem[{{Oh}(2002)}]{oh:02}
{Oh}, S.~P. 2002, \mnras, 336, 1021

\bibitem[{{Oh} \& {Haiman}(2003)}]{oh:03}
{Oh}, S.~P. \& {Haiman}, Z. 2003, \mnras, 346, 456

\bibitem[{{Oh} {et~al.}(2001){Oh}, {Nollett}, {Madau}, \& {Wasserburg}}]{oh:01}
{Oh}, S.~P., {Nollett}, K.~M., {Madau}, P., \& {Wasserburg}, G.~J. 2001, \apjl,
  562, L1

\bibitem[{{Olive} {et~al.}(1987){Olive}, {Thielemann}, \& {Truran}}]{olive:87}
{Olive}, K.~A., {Thielemann}, F., \& {Truran}, J.~W. 1987, \apj, 313, 813

\bibitem[{{O'Meara} {et~al.}(2001){O'Meara}, {Tytler}, {Kirkman}, {Suzuki},
  {Prochaska}, {Lubin}, \& {Wolfe}}]{omeara:01}
{O'Meara}, J.~M., {Tytler}, D., {Kirkman}, D., {Suzuki}, N., {Prochaska},
  J.~X., {Lubin}, D., \& {Wolfe}, A.~M. 2001, \apj, 552, 718

\bibitem[{{Ouchi} {et~al.}(2003){Ouchi}, {Shimasaku}, {Furusawa}, {Miyazaki},
  {Doi}, {Hamabe}, {Hayashino}, {Kimura}, {Kodaira}, {Komiyama}, {Matsuda},
  {Miyazaki}, {Nakata}, {Okamura}, {Sekiguchi}, {Shioya}, {Tamura},
  {Taniguchi}, {Yagi}, \& {Yasuda}}]{ouchi:03}
{Ouchi}, M., {Shimasaku}, K., {Furusawa}, H., {Miyazaki}, M., {Doi}, M.,
  {Hamabe}, M., {Hayashino}, T., {Kimura}, M., {Kodaira}, K., {Komiyama}, Y.,
  {Matsuda}, Y., {Miyazaki}, S., {Nakata}, F., {Okamura}, S., {Sekiguchi}, M.,
  {Shioya}, Y., {Tamura}, H., {Taniguchi}, Y., {Yagi}, M., \& {Yasuda}, N.
  2003, \apj, 582, 60

\bibitem[{{Pascarelle} {et~al.}(1998){Pascarelle}, {Lanzetta}, \& {Fern{\'
  a}ndez-Soto}}]{pascarelle:98}
{Pascarelle}, S.~M., {Lanzetta}, K.~M., \& {Fern{\' a}ndez-Soto}, A. 1998,
  \apjl, 508, L1

\bibitem[{{Pell{\' o}} {et~al.}(2004){Pell{\' o}}, {Schaerer}, {Richard}, {Le
  Borgne}, \& {Kneib}}]{pello:04}
{Pell{\' o}}, R., {Schaerer}, D., {Richard}, J., {Le Borgne}, J.-F., \&
  {Kneib}, J.-P. 2004, \aap, 416, L35

\bibitem[{{Pettini}(2003)}]{pettini:03}
{Pettini}, M. 2003, astro-ph/0303272

\bibitem[{{Pettini} \& {Bowen}(2001)}]{pettini:01}
{Pettini}, M. \& {Bowen}, D.~V. 2001, \apj, 560, 41

\bibitem[{{Pettini} {et~al.}(2002){Pettini}, {Ellison}, {Bergeron}, \&
  {Petitjean}}]{pettini:02}
{Pettini}, M., {Ellison}, S.~L., {Bergeron}, J., \& {Petitjean}, P. 2002, \aap,
  391, 21

\bibitem[{{Prochaska} {et~al.}(2003){Prochaska}, {Gawiser}, {Wolfe}, {Castro},
  \& {Djorgovski}}]{prochaska:03}
{Prochaska}, J.~X., {Gawiser}, E., {Wolfe}, A.~M., {Castro}, S., \&
  {Djorgovski}, S.~G. 2003, \apjl, 595, L9

\bibitem[{{Qian} \& {Wasserburg}(2001)}]{qian:01}
{Qian}, Y.-Z. \& {Wasserburg}, G.~J. 2001, \apj, 559, 925

\bibitem[{{Ricotti}(2002)}]{ricotti:02}
{Ricotti}, M. 2002, \mnras, 336, L33

\bibitem[{{Ricotti} {et~al.}(2004){Ricotti}, {Haehnelt}, {Pettini}, \&
  {Rees}}]{ricotti:04a}
{Ricotti}, M., {Haehnelt}, M.~G., {Pettini}, M., \& {Rees}, M.~J. 2004,
  astro-ph/0403327

\bibitem[{{Ricotti} \& {Ostriker}(2004)}]{ricotti:04b}
{Ricotti}, M. \& {Ostriker}, J.~P. 2004, \mnras, 350, 539

\bibitem[{{Scalo}(1986)}]{scalo:86}
{Scalo}, J.~M. 1986, \fcp, 11, 1

\bibitem[{{Schaerer}(2002)}]{schaerer:02}
{Schaerer}, D. 2002, \aap, 382, 28

\bibitem[{{Schaye} {et~al.}(2003){Schaye}, {Aguirre}, {Kim}, {Theuns}, {Rauch},
  \& {Sargent}}]{schaye:03}
{Schaye}, J., {Aguirre}, A., {Kim}, T., {Theuns}, T., {Rauch}, M., \&
  {Sargent}, W.~L.~W. 2003, \apj, 596, 768

\bibitem[{{Schneider} {et~al.}(2002){Schneider}, {Ferrara}, {Natarajan}, \&
  {Omukai}}]{schneider:02}
{Schneider}, R., {Ferrara}, A., {Natarajan}, P., \& {Omukai}, K. 2002, \apj,
  571, 30

\bibitem[{{Scully} {et~al.}(1997){Scully}, {Casse}, {Olive}, \&
  {Vangioni-Flam}}]{scully:97}
{Scully}, S., {Casse}, M., {Olive}, K.~A., \& {Vangioni-Flam}, E. 1997, \apj,
  476, 521

\bibitem[{{Sembach} {et~al.}(2004){Sembach}, {Wakker}, {Tripp}, {Richter},
  {Kruk}, {Blair}, {Moos}, {Savage}, {Shull}, {York}, {Sonneborn}, {H{\'
  e}brard}, {Ferlet}, {Vidal-Madjar}, {Friedman}, \& {Jenkins}}]{sembach:04}
{Sembach}, K.~R., {Wakker}, B.~P., {Tripp}, T.~M., {Richter}, P., {Kruk},
  J.~W., {Blair}, W.~P., {Moos}, H.~W., {Savage}, B.~D., {Shull}, J.~M.,
  {York}, D.~G., {Sonneborn}, G., {H{\' e}brard}, G., {Ferlet}, R.,
  {Vidal-Madjar}, A., {Friedman}, S.~D., \& {Jenkins}, E.~B. 2004, \apjs, 150,
  387

\bibitem[{{Shapley} {et~al.}(2004){Shapley}, {Erb}, {Pettini}, {Steidel}, \&
  {Adelberger}}]{shapley:04}
{Shapley}, A.~E., {Erb}, D.~K., {Pettini}, M., {Steidel}, C.~C., \&
  {Adelberger}, K.~L. 2004, astro-ph/0405187

\bibitem[{{Simcoe} {et~al.}(2004){Simcoe}, {Sargent}, \& {Rauch}}]{simcoe:04}
{Simcoe}, R.~A., {Sargent}, W.~L.~W., \& {Rauch}, M. 2004, \apj, 606, 92

\bibitem[{{Songaila}(2001)}]{songaila:01}
{Songaila}, A. 2001, \apjl, 561, L153

\bibitem[{{Spergel} {et~al.}(2003){Spergel}, {Verde}, {Peiris}, {Komatsu},
  {Nolta}, {Bennett}, {Halpern}, {Hinshaw}, {Jarosik}, {Kogut}, {Limon},
  {Meyer}, {Page}, {Tucker}, {Weiland}, {Wollack}, \& {Wright}}]{spergel:03}
{Spergel}, D.~N., {Verde}, L., {Peiris}, H.~V., {Komatsu}, E., {Nolta}, M.~R.,
  {Bennett}, C.~L., {Halpern}, M., {Hinshaw}, G., {Jarosik}, N., {Kogut}, A.,
  {Limon}, M., {Meyer}, S.~S., {Page}, L., {Tucker}, G.~S., {Weiland}, J.~L.,
  {Wollack}, E., \& {Wright}, E.~L. 2003, \apjs, 148, 175

\bibitem[{{Steidel} {et~al.}(2001){Steidel}, {Pettini}, \&
  {Adelberger}}]{steidel:01}
{Steidel}, C.~C., {Pettini}, M., \& {Adelberger}, K.~L. 2001, \apj, 546, 665

\bibitem[{{Suda} {et~al.}(2004){Suda}, {Aikawa}, {Machida}, {Fujimoto}, \&
  {Iben Jr}}]{suda:04}
{Suda}, T., {Aikawa}, M., {Machida}, M.~N., {Fujimoto}, M.~Y., \& {Iben Jr}, I.
  2004, astro-ph/0402589

\bibitem[{{Tinsley}(1980)}]{tinsley:80}
{Tinsley}, B.~M. 1980, \fcp, 5, 287

\bibitem[{{Truran} \& {Cameron}(1971)}]{truran:71}
{Truran}, J.~W. \& {Cameron}, A.~G.~W. 1971, \apss, 14, 179

\bibitem[{{Tumlinson} {et~al.}(2004){Tumlinson}, {Venkatesan}, \&
  {Shull}}]{tumlinson:04}
{Tumlinson}, J., {Venkatesan}, A., \& {Shull}, J.~M. 2004, astro-ph/0401376

\bibitem[{{Umeda} \& {Nomoto}(2003)}]{umeda:03}
{Umeda}, H. \& {Nomoto}, K. 2003, \nat, 422, 871

\bibitem[{{Vader}(1986)}]{vader:86}
{Vader}, J.~P. 1986, \apj, 305, 669

\bibitem[{{van den Hoek} \& {Groenewegen}(1997)}]{vandenhoek:97}
{van den Hoek}, L.~B. \& {Groenewegen}, M.~A.~T. 1997, \aaps, 123, 305

\bibitem[{{Venkatesan} {et~al.}(2004){Venkatesan}, {Schneider}, \&
  {Ferrara}}]{venkatesan:04}
{Venkatesan}, A., {Schneider}, R., \& {Ferrara}, A. 2004, \mnras, 349, L43

\bibitem[{{Venkatesan} \& {Truran}(2003)}]{venkatesan:03b}
{Venkatesan}, A. \& {Truran}, J.~W. 2003, \apjl, 594, L1

\bibitem[{{Venkatesan} {et~al.}(2003){Venkatesan}, {Tumlinson}, \&
  {Shull}}]{venkatesan:03a}
{Venkatesan}, A., {Tumlinson}, J., \& {Shull}, J.~M. 2003, \apj, 584, 621

\bibitem[{{Wasserburg} \& {Qian}(2000)}]{wasserburg:00}
{Wasserburg}, G.~J. \& {Qian}, Y.-Z. 2000, \apjl, 538, L99

\bibitem[{{Woosley} \& {Weaver}(1995)}]{woosley:95}
{Woosley}, S.~E. \& {Weaver}, T.~A. 1995, \apjs, 101, 181

\bibitem[{{Wyithe} \& {Loeb}(2003)}]{wyithe:03}
{Wyithe}, J.~S.~B. \& {Loeb}, A. 2003, \apjl, 588, L69

\bibitem[{{Zentner} \& {Bullock}(2003)}]{zentner:03}
{Zentner}, A.~R. \& {Bullock}, J.~S. 2003, \apj, 598, 49

\end{thebibliography}

\end{document}